\documentclass[%
 reprint,
showkeys,
 amsmath,amssymb,
 aps,
Prb,
]{revtex4-2}

\usepackage{graphicx}
\usepackage{dcolumn}
\usepackage{bm}
\usepackage{color}
\usepackage{subfigure}

\begin{document}

\preprint{APS/123-QED}

\title{Harnessing $\mathcal{PT}$-symmetry in non-Hermitian stiffness-modulated waveguides}

\author{Emanuele Riva}
 \affiliation{Department of Mechanical Engineering, Politecnico di Milano, Milano 20156, Italy}

\date{\today}

\begin{abstract}
The recent progress in the context of elastic metamaterials and modulated waveguides with digitally controllable properties has opened new pathways to overcome the limitations dictated by Hermitian Hamiltonians in mechanics. Among the possible implementations, non-Hermitian, $\mathcal{PT}$-symmetric systems with balanced gain and loss have emerged as an elegant mechanism to access novel functionalities by lifting the non-Hermitian degeneracies (exceptional points).
Motivated by this, the paper deals with a non-Hermitian and $\mathcal{PT}$-symmetric elastic waveguide with complex stiffness-modulation. 
The strength of the stiffness-modulations, tailored in the form of a balanced gain/loss, delineates a transition from unbroken to broken $\mathcal{PT}$-symmetric phases, where distinct Bloch-wave modes coalesce into exceptional points. It is shown that, in the unbroken $\mathcal{PT}$-symmetric regime, and due to the interplay between real and imaginary components of the elasticity, the waveguide operates as a phononic filter. 
When the strength of the gain/loss interactions increases, the frequency gap closes and the bulk bands degenerate into an exceptional point, where the system operates as a waveguide with asymmetric scattering capabilities.
The paper provides a connection between the distinct wave modes that populate the non-Hermitian degeneracies and the directional reflection/transmission capabilities. The asymmetric behavior is herein explained by combining the dispersion properties of a $\mathcal{PT}$-symmetric rod, obtained through the plane wave expansion method (PWEM), and the scattering matrix method (SMM) for a modulated slab series-connected to semi infinite media.
\end{abstract}


\maketitle


\section*{Introduction}
Modulated materials are key for many desired dynamic properties in the context of phononic structures and can be regarded as building blocks for the implementation of complex systems with unusual functionalities. Among the relevant works, topological waveguides \cite{miniaci2018experimental,susstrunk2015observation,vila2017observation,riva2018tunable,liu2018tunable,rosa2019edge,riva2020edge}, cloaking \cite{norris2008acoustic,quadrelli2021elastic,QUADRELLI2021116396,chen2015latticed}, rainbow devices \cite{PhysRevApplied.16.034028,de2020experimental,alshaqaq2020graded,chaplain2020topological}, lenses \cite{tol2017phononic}, should be mentioned. These examples offer non-conventional energy transfer mechanisms and rely on passive geometries, stiffness modulations, and symmetry breaks, which have been sought in analogy to similar behaviors previously observed in different realms of physics. However, the search of new functionalities, such as nonreciprocity \cite{marconi2020experimental,trainiti2016non,attarzadeh2019non}, digitally controllable waveguides \cite{xia2021experimental,riva2021adiabatic}, frequency conversion \cite{yi2018one,yi2017frequency}, and parametric amplification \cite{trainiti2019time}, requires more advanced space-time control of the material parameters, which justifies the emergence of active times in phononics \cite{zangeneh2019active}. 
The experimental works done in this direction have \textit{de facto} opened new opportunities to engineer complex media with exotic properties and have spurred additional research on this matter. An emerging field, very active in other physical realms but less explored especially in experimental mechanics, which can take advantage of this new take on modulated systems, relies on non-Hermitian systems with $\mathcal{PT}$-symmetry \cite{longhi2018parity,feng2017non,longhi2009bloch}. Despite non-Hermitian systems in many cases are characterized by complex spectra and non-unitary evolution of the states, there are special symmetries where a fine tuning of gain/loss interactions yields phases with purely real spectra (unbroken $\mathcal{PT}$-symmetric phase), and complex eigenstates (broken $\mathcal{PT}$-symmetric phase), separated by spectral singularities, known as exceptional points (EP) \cite{bender1998real,bender2019pt,longhi2010spectral}. In other words, when non-Hermiticity marries $\mathcal{PT}$-symmetry, the hypothesis of Hermiticity is replaced by $\mathcal{PT}$-symmetric invariance, and additional functionalities can be accessed. Motivated by this, the manuscript emulates asymmetric scattering in the context of longitudinal waves in 1D solid, which is pursued through a waveguide with complex sinusoidal elasticity that not only serves to induce Bragg-scattering mechanisms, but also to activate balanced gain and loss interactions. 
In contrast to prior works on the topic \cite{fleury2015invisible,PhysRevLett.119.243904,wu2019asymmetric}, the plane wave expansion method (PWEM) \cite{riva2019generalized,enrico2020omindirectional,trainiti2016non} is used to show that there is a synergistic interplay between the modulation strengths of the real and imaginary parts of the stiffness profile, which coalesce the dispersion bands and give birth to EPs. 
The modulation parameters are linked to the unbroken and broken $\mathcal{PT}$-symmetric phases, to provide a quantitative analysis of the phase transition. It is shown that, despite sharing the same frequency, left and right propagating wave-modes are inherently different due to the complex interplay between real and imaginary parts of the stiffness. The asymmetric wave modes are combined to the scattering matrix method (SMM) \cite{yi2018one} to justify the emergence of asymmetric reflections induced by the modulation, and to qualify the scattering properties of the PT-symmetric slab when series-connected to two semi-infinite passive media. 
\section{Dispersion analysis}
\begin{figure*}[t!]
	\centering
	\subfigure[]{\includegraphics[width=0.35\textwidth]{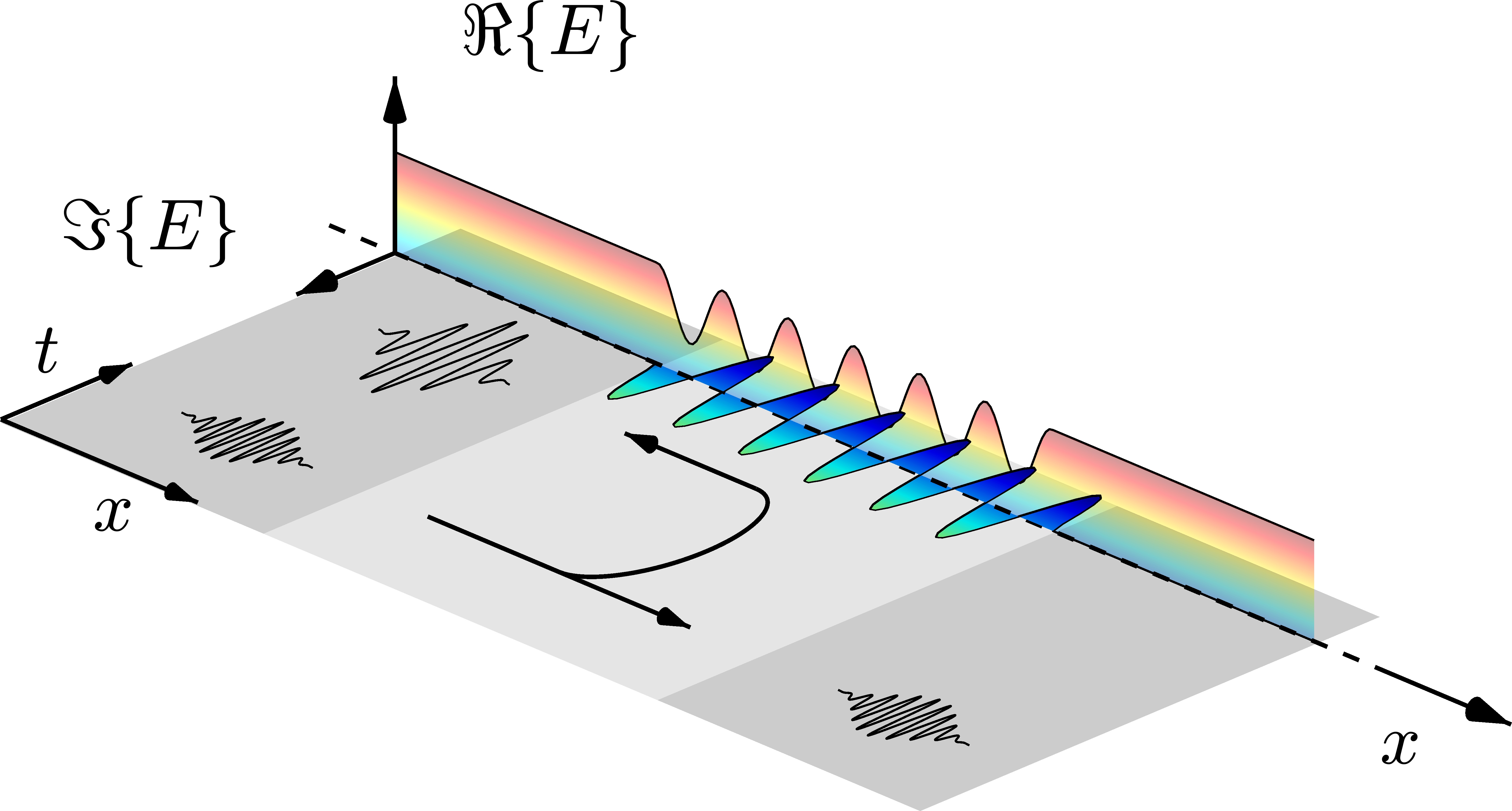}}\hspace{1cm}
	\subfigure[]{\includegraphics[width=0.35\textwidth]{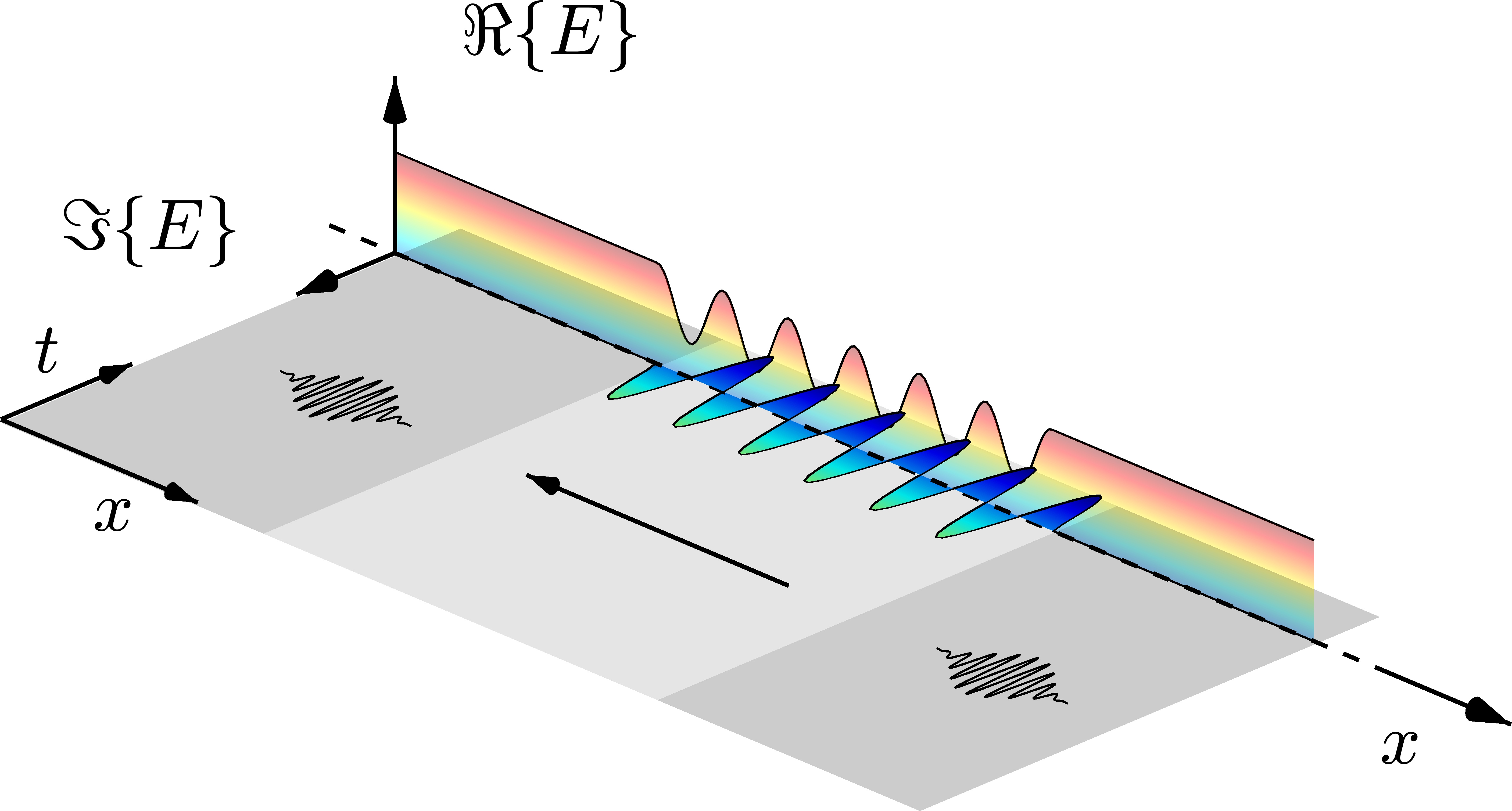}}\\
	\subfigure[]{\includegraphics[width=0.32\textwidth]{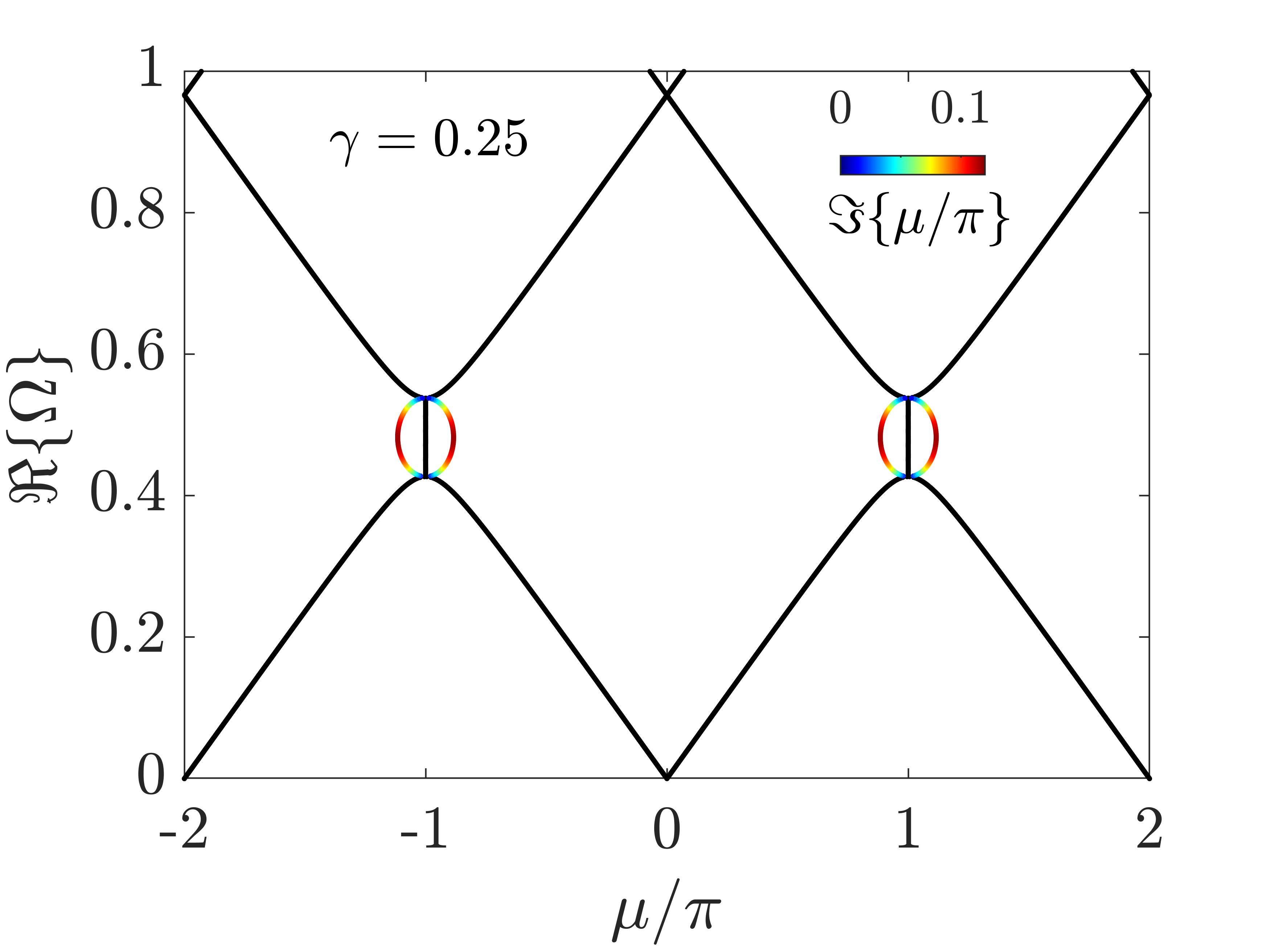}}	
	\subfigure[]{\includegraphics[width=0.32\textwidth]{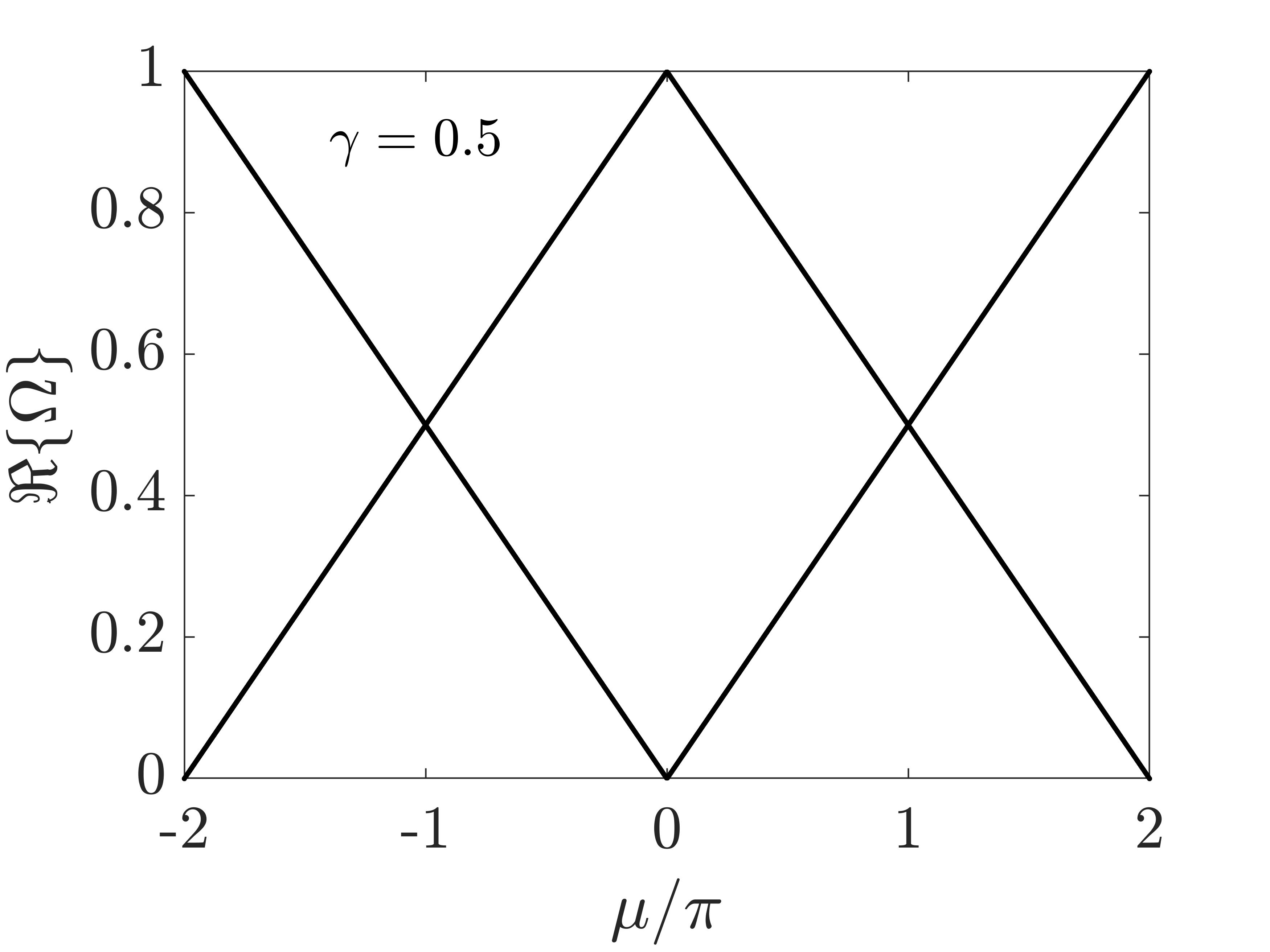}}	
	\subfigure[]{\includegraphics[width=0.32\textwidth]{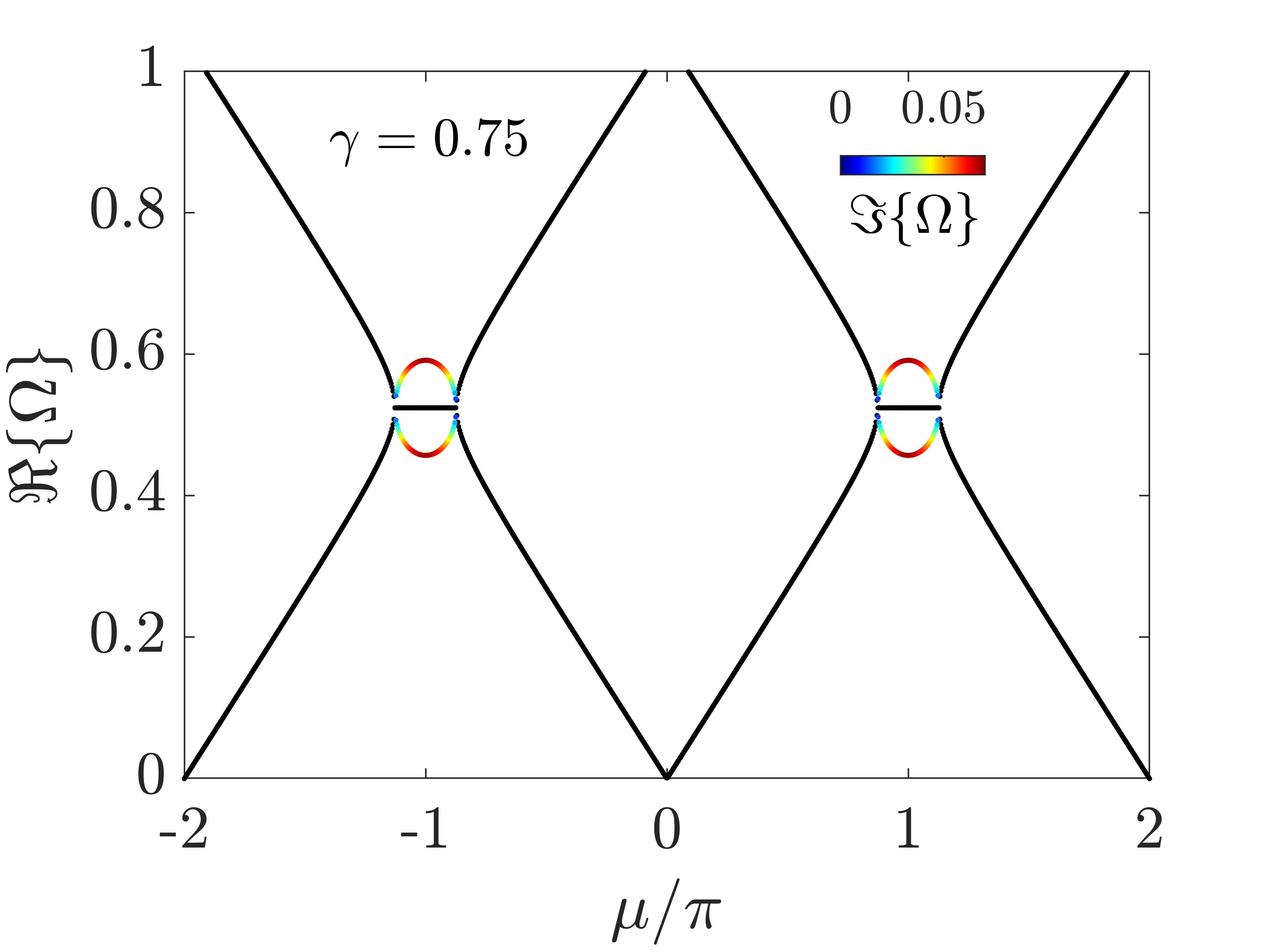}}\\
	\subfigure[]{\includegraphics[width=0.24\textwidth]{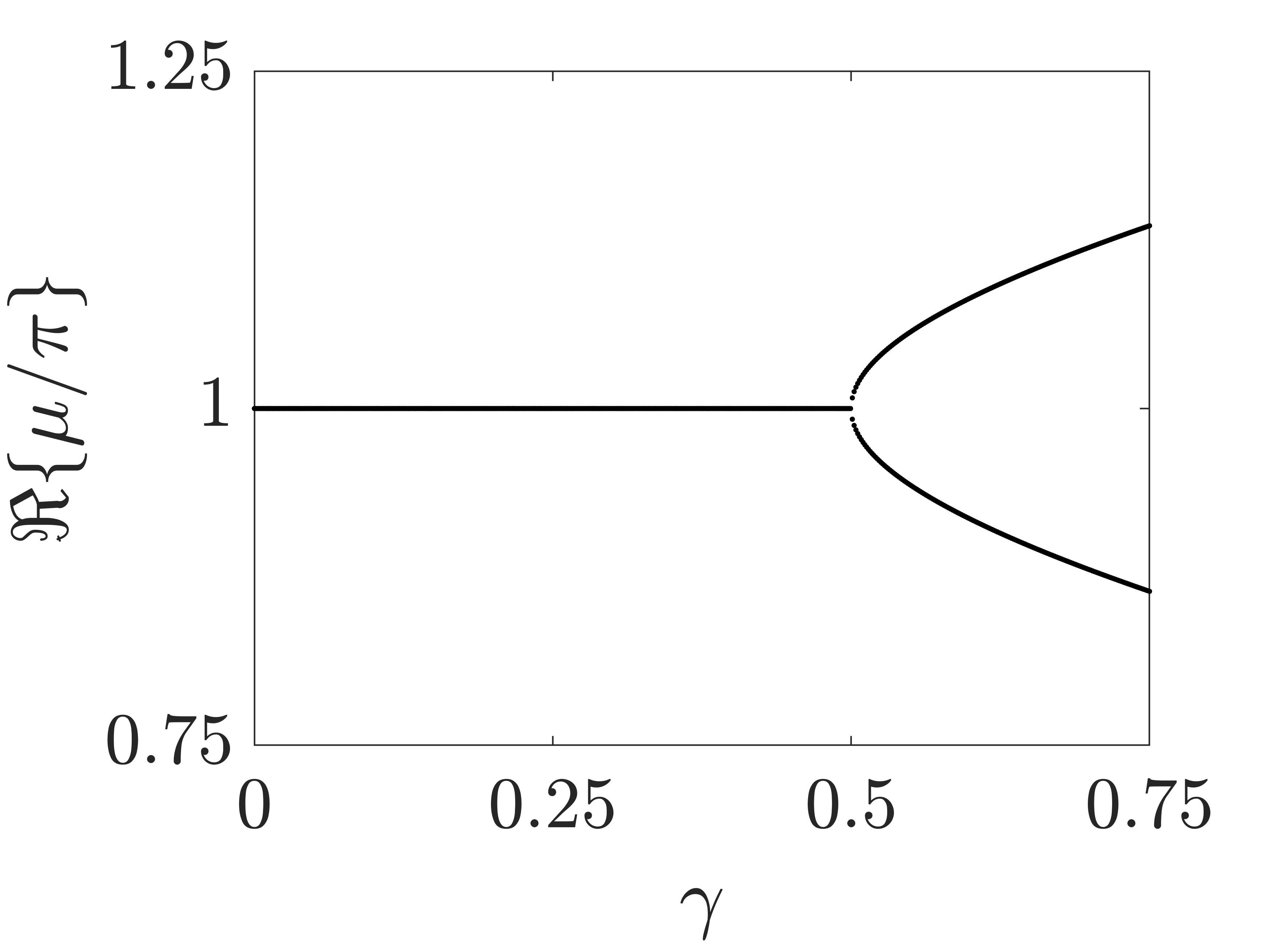}}	
	\subfigure[]{\includegraphics[width=0.24\textwidth]{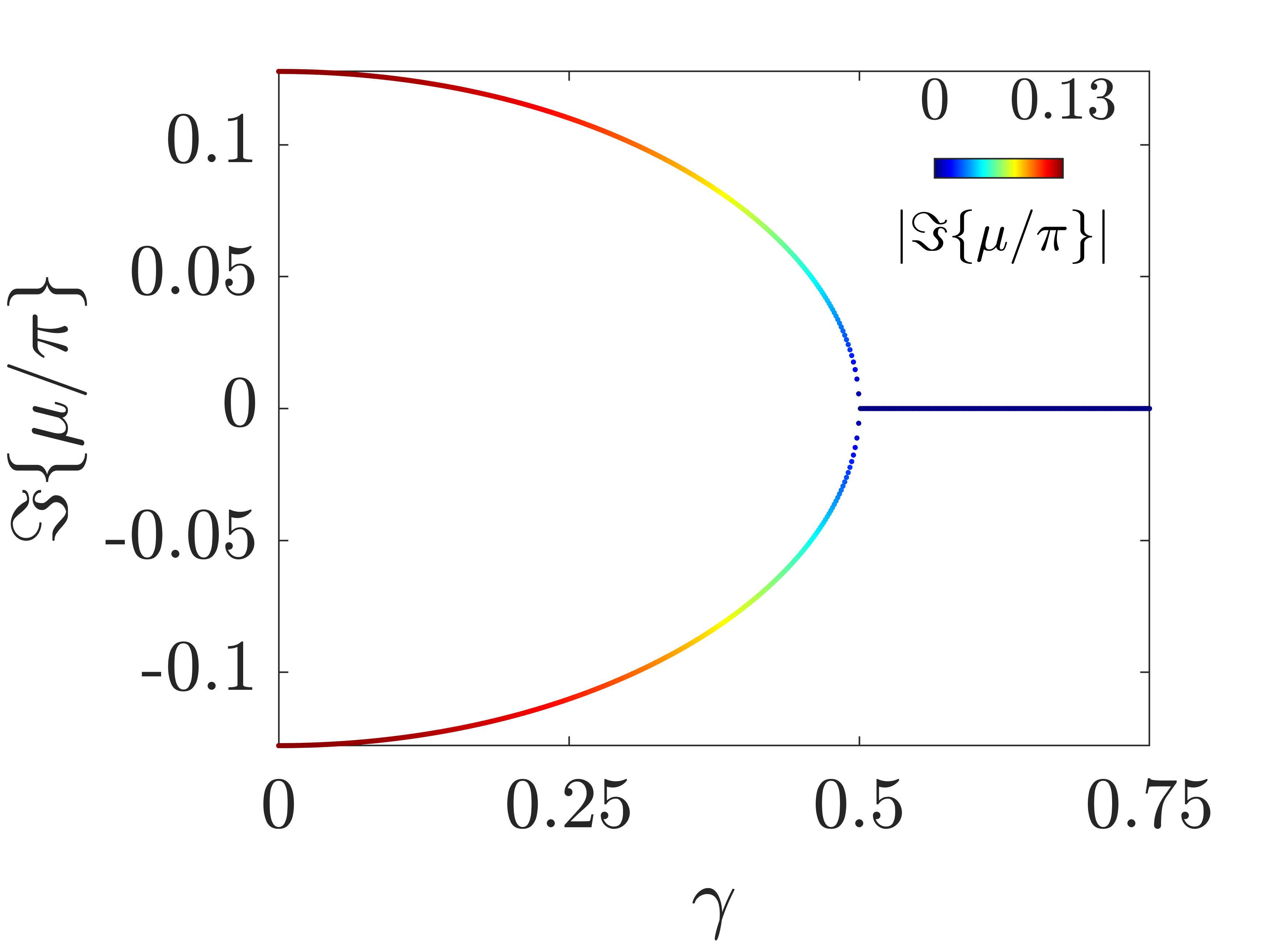}}	
	\subfigure[]{\includegraphics[width=0.24\textwidth]{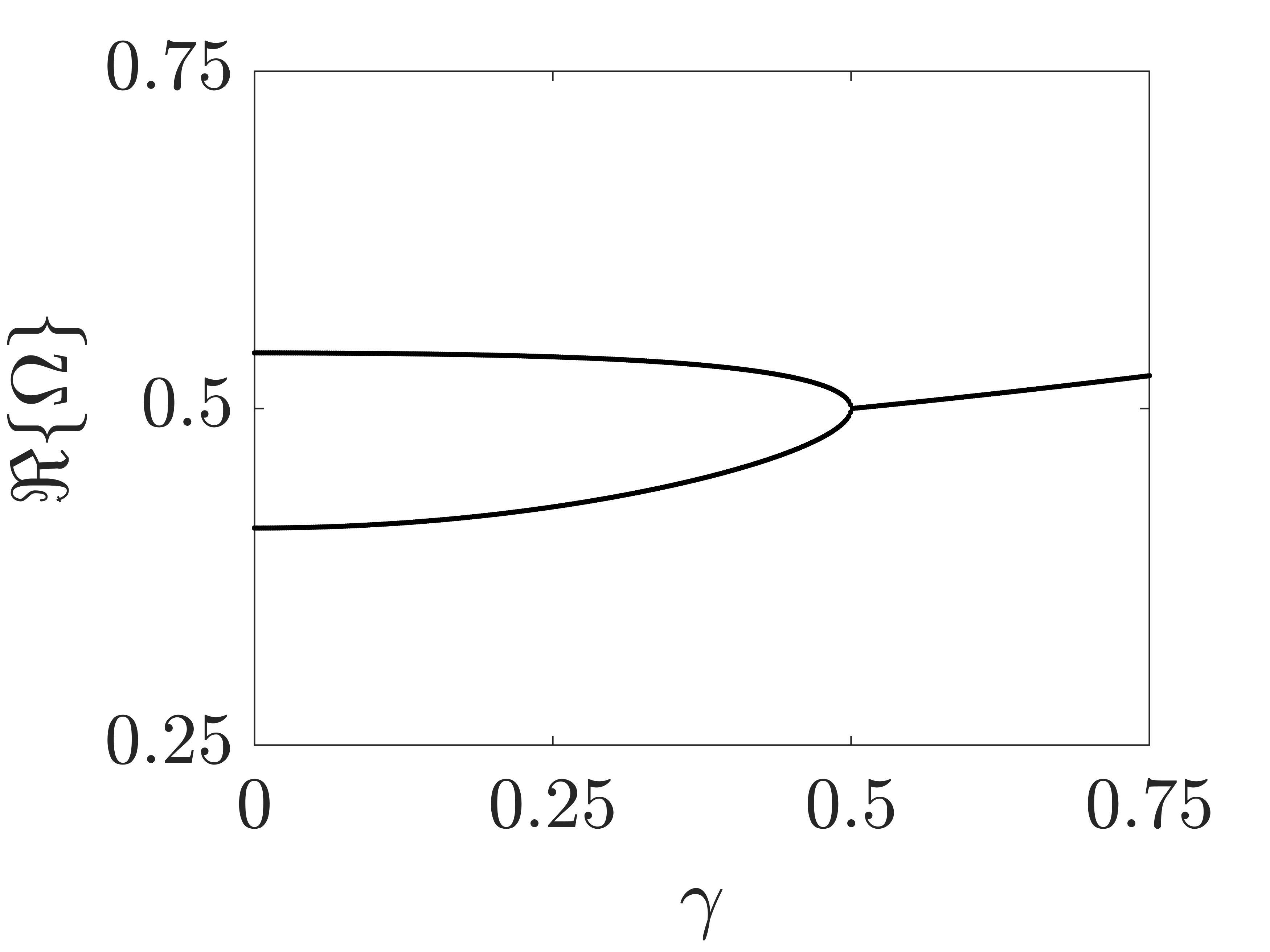}}	
	\subfigure[]{\includegraphics[width=0.24\textwidth]{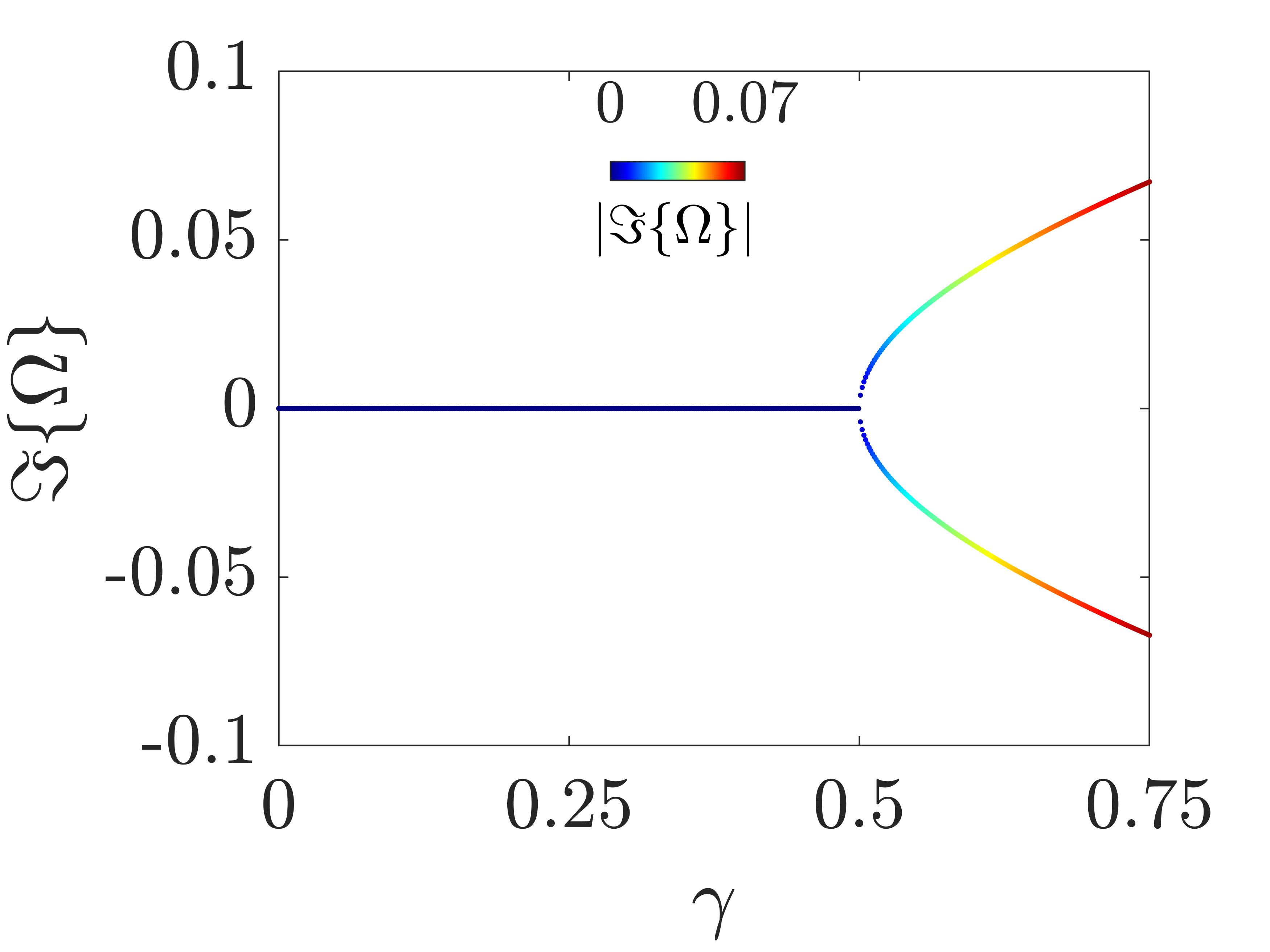}}	
	\caption{(a-b) Schematic of the modulated slab and concept. The real and imaginary components of the complex modulation are projected onto perpendicular planes and are represented by colored surfaces. (a) A wave incident from the left is reflected by the slab with amplification and transmitted with unitary gain. (b) A wave incident from the right is transmitted unaltered without reflection. (c-e) Dispersion relation  (c) $\mu\left(\Omega\right)$ for modulation parameters $\alpha=0.5$ and $\gamma=0.25$, (d) $\alpha=\gamma=0.5$, and (e) $\Omega\left(\mu\right)$ for $\alpha=0.5$ and $\gamma=0.75$. (f-g) Wavenumbers evaluated at constant frequency ($\Omega=0.5$) upon varying $\gamma$. There is a complex wavenumber split at $\gamma=0.5$ for negative variations of $\gamma$. (h-i) Frequency transition evaluated at constant wavenumber ($\mu/\pi=1$) upon varying $\gamma$. There is a complex frequency split at $\gamma=0.5$ for positive variations of $\gamma$.  }
	\label{Fig1}    
\end{figure*}
Consider the waveguide illustrated in Figure \ref{Fig1}(a-b). As mentioned, aim of the paper is to discuss the directional scattering capabilities of the elastic slab and to provide a clear connection with the dispersion properties. The concept is elucidated in the figure: a wave impinging on the modulation from the left is reflected and, eventually, amplified. While a wave impinging from the opposite side propagates with unitary amplitude and with no reflection.
The study is initially focused on the dispersion properties of the modulated part, represented with the colored surfaces, which embodies a complex elasticity $E\left(x\right)=E_0\left[1+\alpha\cos{\left(\kappa_mx\right)}\right]+\rm{i}E_0\gamma\sin{\left(\kappa_mx\right)}$ having modulation wavenumber $\kappa_m=2\pi/\lambda_m$ and wavelength $\lambda_m$. $\alpha$ and $\gamma$ are the modulation strengths. The former is assumed fixed ($\alpha=0.5$) without loss of generality, while the role of $\gamma$ is object of discussion. Note now that the function is even in its real part and the imaginary part is odd. As such, $E\left(x\right)$ is invariant under the combination of reflection $x\rightarrow-x$ and time reversal $\rm{i}\rightarrow-\rm{i}$ operators. 
Longitudinal wave motion $u\left(x,t\right)$ in such a system is governed by the following elastodynamic equation:
\begin{equation}
    \left(E\left(x\right)u_{,x}\right)_{,x}=\left(\rho u_{,t}\right)_{,t}
    \label{eq:01}
\end{equation}
where $(\cdot)_{,\xi}$ stands for $\partial(\cdot)/\partial\xi$ and, without loss of generality, the material density $\rho$ is assumed constant. 
Due to the periodic nature of the underlying medium, i.e. $E\left(x\right)=E\left(x+\lambda_m\right)$, the dispersion analysis is encouraged and hereafter accomplished by following the plane wave expansion method (PWEM) \cite{trainiti2016non,riva2019generalized}. As such, the elasticity is expanded in terms of $2N+1$ complex exponential functions $E\left(x\right)=\sum_{n=-N}^{N}\hat{E}_{n}\rm{e}^{-\rm{i}n\kappa_mx}$, where $\hat{E}_{n}$ are the Fourier coefficients and $N=10$ is found sufficient to describe the dispersion properties with good approximation. Ansatz $u\left(x,t\right)$ are in the same complex exponential form $u\left(x,t\right)= {\rm{u}}\left(x\right)\rm{e}^{\rm{i}\left(\omega t-\kappa x\right)}$, with ${\rm{u}}\left(x\right)=\sum_{p=-P}^{P}\hat{u}_{p}\rm{e}^{-\rm{i}p\kappa_mx}$. These are combined to Eq. \ref{eq:01} and the orthogonality of the exponential functions is exploited to get to the dispersion relation in the form of a quadratic eigenvalue problems for $\omega\left(\kappa\right)$ and $\kappa\left(\omega\right)$:
\begin{equation}
\sum_{p=-P}^P\hat{E}_{s-p}\left(\kappa+p\kappa_m\right)\left(\kappa+s\kappa_m\right)\hat{u}_{p}=\omega^2\hat{u}_{s}
\label{eq:02}
\end{equation}
Now, the dispersion relation in dimensionless form is illustrated in Figures \ref{Fig1}(b-d) for fixed $\alpha=0.5$ and for three distinct $\gamma$ values (i.e. $\gamma=0.25$, $\gamma=0.5$, and $\gamma=0.75$), where the dimensionless frequency and wavenumber are $\Omega=\omega\lambda_m/c_0$ and $\mu=\kappa\lambda_m$ with $c_0=\sqrt{E_0/\rho}$.
Interestingly, the configuration with $\gamma<\alpha$ (see Figure \ref{Fig1}(c)) displays real frequencies and complex wavenumbers, induced by bandgap formation mechanisms at the lattice points. Such a frequency gap is reduced in width until it coalesces to a non-Hermitian degeneracy in case $\gamma=\alpha$. Under such a condition, displayed in Figure \ref{Fig1}(d), the frequency-wavenumber pairs are purely real. When the modulation strength $\gamma$ is further increased to $\gamma>\alpha$, there is a complex frequency split and a $\kappa$-gap is induced in the neighborhood of the lattice points, see Figure \ref{Fig1}(e). To gain insight on the physics of phase transition, Eq. \ref{eq:02} is truncated to the first term, i.e. $N=1$ and the analytical form of the Fourier coefficients is used. The eigenvalue problem is conveniently rewritten in matrix form $H\left(\kappa\right)\bm{u}=\rho\omega^2I\bm{u}$, where $\bm{u}=\left[u_{-1},u_0,u_1\right]^T$ is the eigenvector that accommodates the $\hat{u}_p$ Bloch-wave coefficients of the expansion, and the matrix $H$ reads:
\begingroup\makeatletter\def\f@size{9}\check@mathfonts
\begin{equation}
\begin{split}
    H=&\;E_0\begin{bmatrix}
    \left(\kappa-\kappa_m\right)^2&\displaystyle\frac{\kappa}{2}\alpha\left(\kappa-\kappa_m\right)&0\\[5pt]
    \displaystyle\frac{\kappa}{2}\alpha\left(\kappa-\kappa_m\right)&\kappa^2&\displaystyle\frac{\kappa}{2}\alpha\left(\kappa+\kappa_m\right)\\[5pt]
    0&\displaystyle\frac{\kappa}{2}\alpha\left(\kappa+\kappa_m\right)&\left(\kappa+\kappa_m\right)^2
    \end{bmatrix}+\\[5pt]
    &\;E_0\begin{bmatrix}
    0&\displaystyle\frac{\kappa}{2}\gamma\left(\kappa-\kappa_m\right)&0\\[5pt]
    -\displaystyle\frac{\kappa}{2}\gamma\left(\kappa-\kappa_m\right)&0&\displaystyle\frac{\kappa}{2}\gamma\left(\kappa+\kappa_m\right)\\[5pt]
    0&-\displaystyle\frac{\kappa}{2}\gamma\left(\kappa+\kappa_m\right)&0
    \end{bmatrix}
\end{split}
\end{equation}
\endgroup
here, the contributions due to $\alpha$ and $\gamma$ are separated. It is observed that $\alpha$ is associated with the Hermitian part of the matrix $H$, while $\gamma$ induces non-Hermiticity. Also, due to the aforementioned considerations for $E\left(x\right)$, $H$ commutes with the $\mathcal{PT}$ operator, i.e., $\left(H,\mathcal{PT}\right)=0$, which suggests the emergence of a phase transition between unbroken and broken $\mathcal{PT}$-symmetric phases. This is further confirmed by spanning the parameter space $\gamma$ for the fixed $\alpha$ value, as shown in figure \ref{Fig1}(f-i) and object of discussion. Figure \ref{Fig1}(f-g) illustrate a transition for a fixed frequency $\Omega=0.5$, in which complex-conjugate wavenumber pairs split for negative variations of $\gamma$. Hence, the dispersion relation is sought in the form $\mu\left(\Omega\right)$ to show that $\gamma<\alpha$ yields the formation of a frequency gap, while in case $\gamma>\alpha$ the wavenumbers are purely real and the dispersion is to be sought in the form $\Omega\left(\mu\right)$. In contrast, the behavior for $\mu/\pi=1$, shown in figure \ref{Fig1}(h-i), is characterized by frequencies initially real which coalesce if $\gamma=\alpha$ and, due to a further increase of $\gamma$ split into complex conjugate pairs. \\
To conclude the dispersion analysis of the slab, the discussion is now focused on the behavior at the non-Hermitian degeneracy, which is preparatory to the following. To that end, $\gamma=\alpha$ is imposed into Eq. \ref{eq:02} with $N=1$ harmonics and the characteristic equation relative to the eigenvalue problem in dimensionless form $\left(\Omega^2I-H\right)\bm{u}=0$ is:
\begin{equation}
\begin{split}
    det\;\left(\Omega^2I-H\right)=\left[\Omega^2-\left(\frac{\mu^2}{4\pi^2}-\frac{\mu}{\pi}+1\right)\right]\left[\Omega^2+\right.\\[5pt]
    -\left.\frac{\mu^2}{4\pi^2}\right]\left[\Omega^2-\left(\frac{\mu^2}{4\pi^2}+\frac{\mu}{\pi}+1\right)\right]
\label{eq:det}
\end{split}
\end{equation}
by looking for coincident solutions of Eq. \ref{eq:det} it is here demonstrated that, independently of frequency $\Omega$, the eigenvalues coalesce in the origin $\mu=0$ and, more interestingly, for $\mu=\pm\pi$. Now, plugging $\mu=\pi$ (i.e. forward-traveling waves) into Eq. \ref{eq:02} gives the frequencies of the non-Hermitian degeneracy $\Omega=\pm\frac{1}{2}$ both with algebraic multiplicity 2 and $\Omega=\pm\frac{3}{2}$ with algebraic multiplicity 1. This latter is not relevant for the scope of the present paper and corresponds to a higher frequency crossing. Notice that $\mu=-\pi$ also gives $\Omega=\pm\frac{1}{2}$ and $\Omega=\pm\frac{3}{2}$.
The study is focused on counter-propagating waves at EPs, i.e. when $\Omega=\frac{1}{2}$ and $\mu=\pm\pi$ and the matrix $H$ is rank deficient, since here the geometric multiplicity is 1. Interestingly, even though opposite wavenumbers share the same frequency, the eigenvectors are different, which suggests that directional wave motion supported by the slab:
\begin{equation}
\begin{split}
    \bm{u}^+&=\begin{pmatrix}
    1\\[5pt]
    0\\[5pt]
    0
    \end{pmatrix}\;\;\;for\;\;\mu^+=\pi;\\[5pt]
    \bm{u}^-&=\begin{pmatrix}
    -\displaystyle\frac{3}{8}\alpha\\[7pt]
    1\\[3pt]
    0
    \end{pmatrix}\;\;\;for\;\;\mu^-=-\pi;
\end{split}
\label{eq:05}
\end{equation}
This is key for the following discussion.
\section{Scattering analysis}
\begin{figure*}[t!]
	\centering
	\subfigure[]{\includegraphics[width=0.43\textwidth]{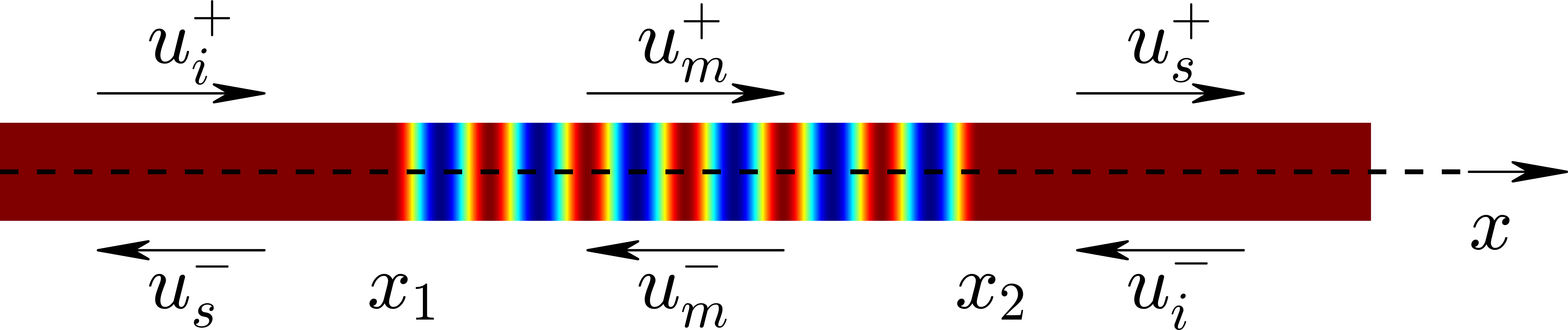}}\\
	\subfigure[]{\includegraphics[width=0.48\textwidth]{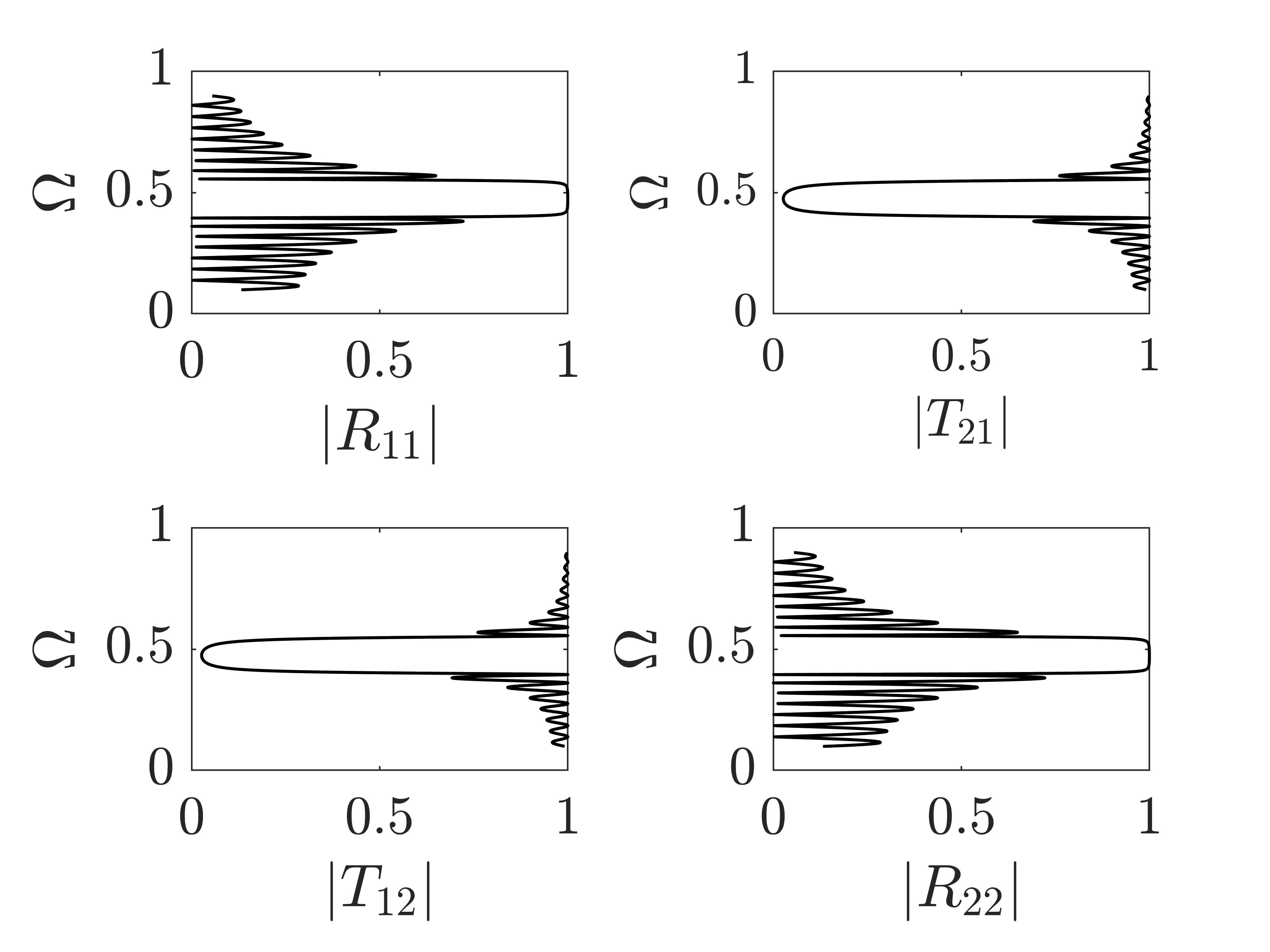}}
	\subfigure[]{\includegraphics[width=0.48\textwidth]{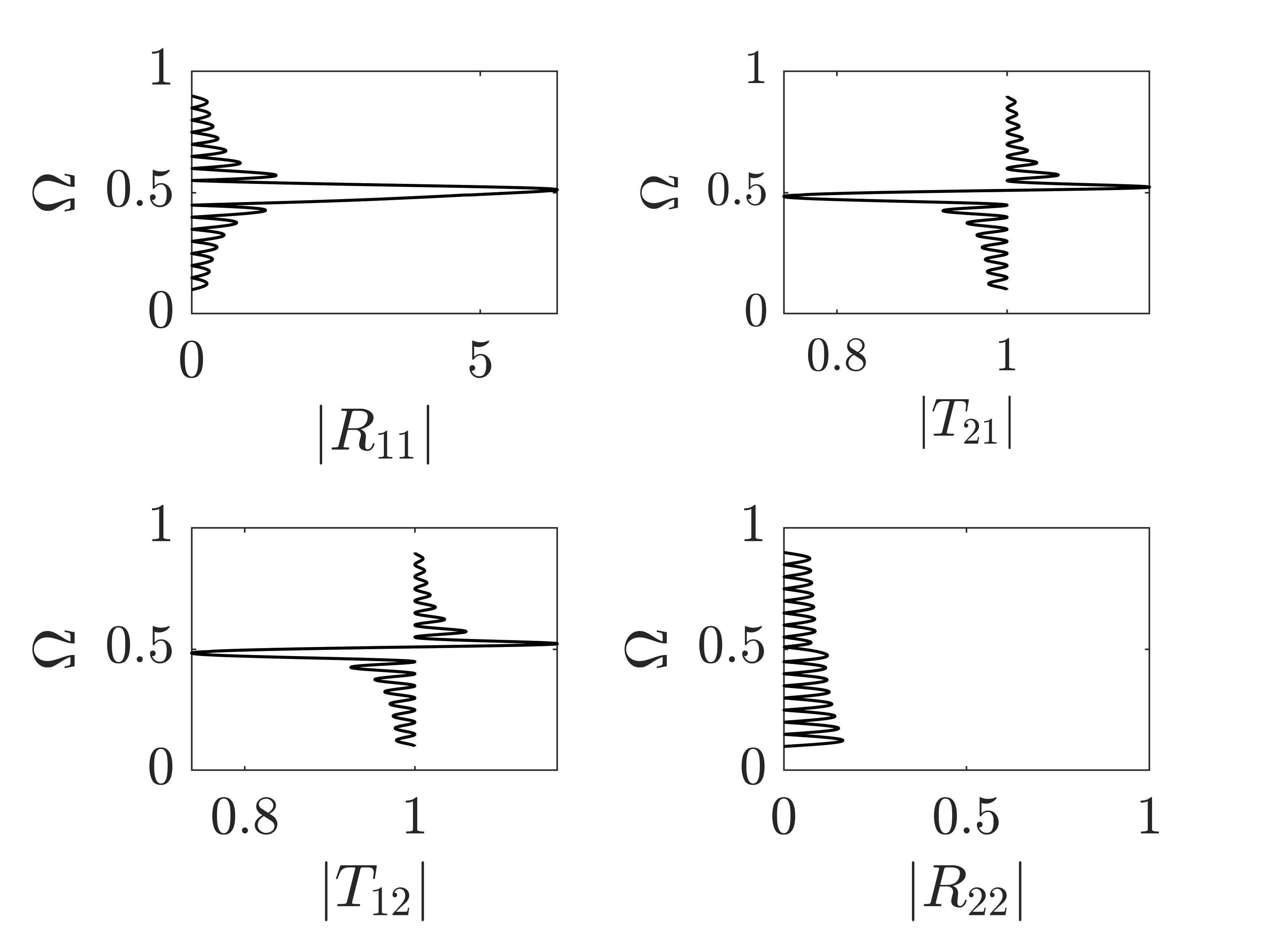}}
	\caption{(a) Schematic of the modulated slab (colored region) connected to the left and to the right with homogeneous semi-infinite elements (red domains). In the figure, the incident and scattered waves are represented with arrows. Scattering coefficients upon varying frequency $\Omega$ for  (b) $\alpha=0.5$ and $\gamma=0$, and for (c) $\alpha=\gamma=0.5$. }
	\label{Fig2}    
\end{figure*}
Consider now a slab made of $Q=10$ unit cells, which is connected to two semi-infinite and homogeneous beam elements with stiffness $E_0\left(1+\alpha\right)$, $c_{0,\alpha}=\sqrt{E_0\left(1+\alpha\right)/\rho}$, and length $L_e=20 \lambda_m$. Note that an integer number of unit cells is required for the modulated rod to embody balanced gain/loss interactions. 
The reference system is arbitrarily set according to the schematic in Figure \ref{Fig2}(a) and, in the following, a number of incident and scattered waves are considered. 
That is, a plane wave $u_i^+$ that impinges on the modulated slab from the left, and a plane wave $u_i^-$ that impinges from the right are input of the problem:
\begin{equation}
    u_i^+=A_0{\rm e}^{{\rm i}\displaystyle\omega\left(-\displaystyle\frac{x}{c_{0,\alpha}}+t\right)}\hspace{0.5cm}
    u_i^-=G_0{\rm e}^{{\rm i}\displaystyle\omega\left(\displaystyle\frac{x}{c_{0,\alpha}}+t\right)}
    \label{eq:06}
\end{equation}
due to the interface conditions between the homogeneous and the modulated media, a pair of scattered waves $u_s^+$ and $u_s^-$ populate the region prior to and after the slab:
\begin{equation}
    u_s^+=F_0{\rm e}^{{\rm i}\displaystyle\omega\left(-\displaystyle\frac{x}{c_{0,\alpha}}+t\right)}\hspace{0.5cm}
    u_s^-=B_0{\rm e}^{{\rm i}\displaystyle\omega\left(\displaystyle\frac{x}{c_{0,\alpha}}+t\right)}
    \label{eq:07}
\end{equation}
finally, the modulated medium is populated by left and right propagating terms $u_m^+$ and $u_m^-$:
\begin{equation}
\begin{split}
    u_m^+=C_0\sum_{p=-P}^P\hat{u}_{p,0}^+{\rm e}^{\displaystyle{\rm -i}\left(\kappa_0^++p\kappa_m\right)x}{\rm e}^{\displaystyle{\rm i}\omega t}\\[5pt]
    u_m^-=D_0\sum_{p=-P}^P\hat{u}_{p,0}^-{\rm e}^{\displaystyle{\rm -i}\left(\kappa_0^-+p\kappa_m\right)x}{\rm e}^{\displaystyle{\rm i}\omega t}    
\end{split}
\label{eq:08}
\end{equation}
where the subscript $0$ stands for central branch of the dispersion in the form $\kappa\left(\omega\right)$ and the superscripts $\pm$ stands for right and left right traveling waves with wavenumbers $\kappa_0^{\pm}$, which are identified and separated by way of the group velocity $c_g=\omega/\kappa_0$ (see the supplementary material for more details on this procedure \cite{SM}). 
The analysis is continued by enforcing compatibility of the displacements and force equilibrium at the interface coordinates $x_1$ and $x_2$. These conditions are hereafter employed to find a relationship between the amplitude coefficients defined in Eq. \ref{eq:06}-\ref{eq:08}. For the boundary at $x=x_1$:
\begin{equation}
\begin{split}
    &u_i^+\left(x_1\right)+u_s^-\left(x_1\right)=u_m^+\left(x_1\right)+u_m^-\left(x_1\right)\\[5pt]
    &E_0\left(1+\alpha\right)\left(u_i^+\left(x_1\right)+u_s^-\left(x_1\right)\right)_{,x}=\\
    &\hspace{2.5cm} E\left(x_1\right)\left(u_m^+\left(x_1\right)+u_m^-\left(x_1\right)\right)_{,x}
\end{split}
\label{eq:09}
\end{equation}
and for $x=x_2$:
\begin{equation}
\begin{split}
    &u_m^+\left(x_2\right)+u_m^-\left(x_2\right)=u_i^-\left(x_2\right)+u_s^+\left(x_2\right)\\[5pt]
    &E\left(x_2\right)\left(u_m^+\left(x_2\right)+u_m^-\left(x_2\right)\right)_{,x}=\\
    &\hspace{2.5cm}E_0\left(1+\alpha\right)\left(u_i^-\left(x_2\right)+u_s^+\left(x_2\right)\right)_{,x}
\end{split}
\label{eq:10}
\end{equation}
notice that the wave modes, reported into Eq. \ref{eq:05}, are asymmetric at the exceptional points and take part in the scattering process described in Eq. \ref{eq:09}-\ref{eq:10}. It is therefore expected that asymmetric scattering emerges for frequencies where the wave modes are asymmetric. The scattering matrix $S$, which links the incident and the scattered wave amplitudes, is obtained after a few manipulations of Eq. \ref{eq:09}-\ref{eq:10} (see the supplementary material \cite{SM} for additional details on the derivation):
\begin{equation}
    \begin{pmatrix}
    B_0\\[5pt]
    F_0
    \end{pmatrix}=\begin{bmatrix}
    R_{11}&T_{21}\\[5pt]
    T_{12}&R_{22}
    \end{bmatrix}\begin{pmatrix}
    A_0\\[5pt]
    G_0
    \end{pmatrix}=S\begin{pmatrix}
    A_0\\[5pt]
    G_0
    \end{pmatrix}
    \label{eq:scattering}
\end{equation}
The coefficients $R_{11}$, $T_{12}$, $T_{21}$, and $R_{22}$ are displayed in figures \ref{Fig2}(b,c) for $\gamma=0$ and $\gamma=\alpha$, respectively. 
\begin{figure*}[t!]
	\centering
	\subfigure[]{\includegraphics[width=0.4\textwidth]{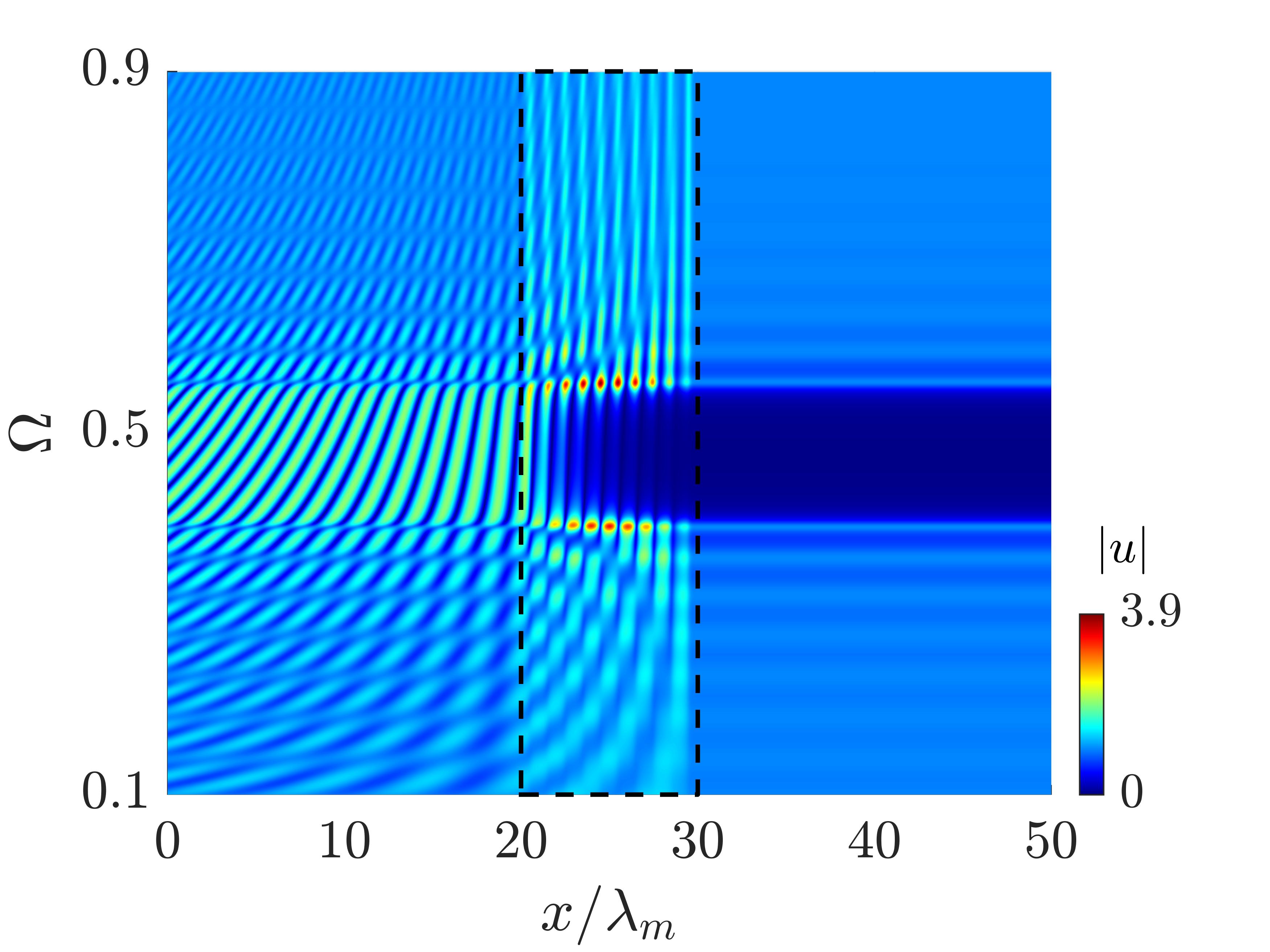}}
	\subfigure[]{\includegraphics[width=0.4\textwidth]{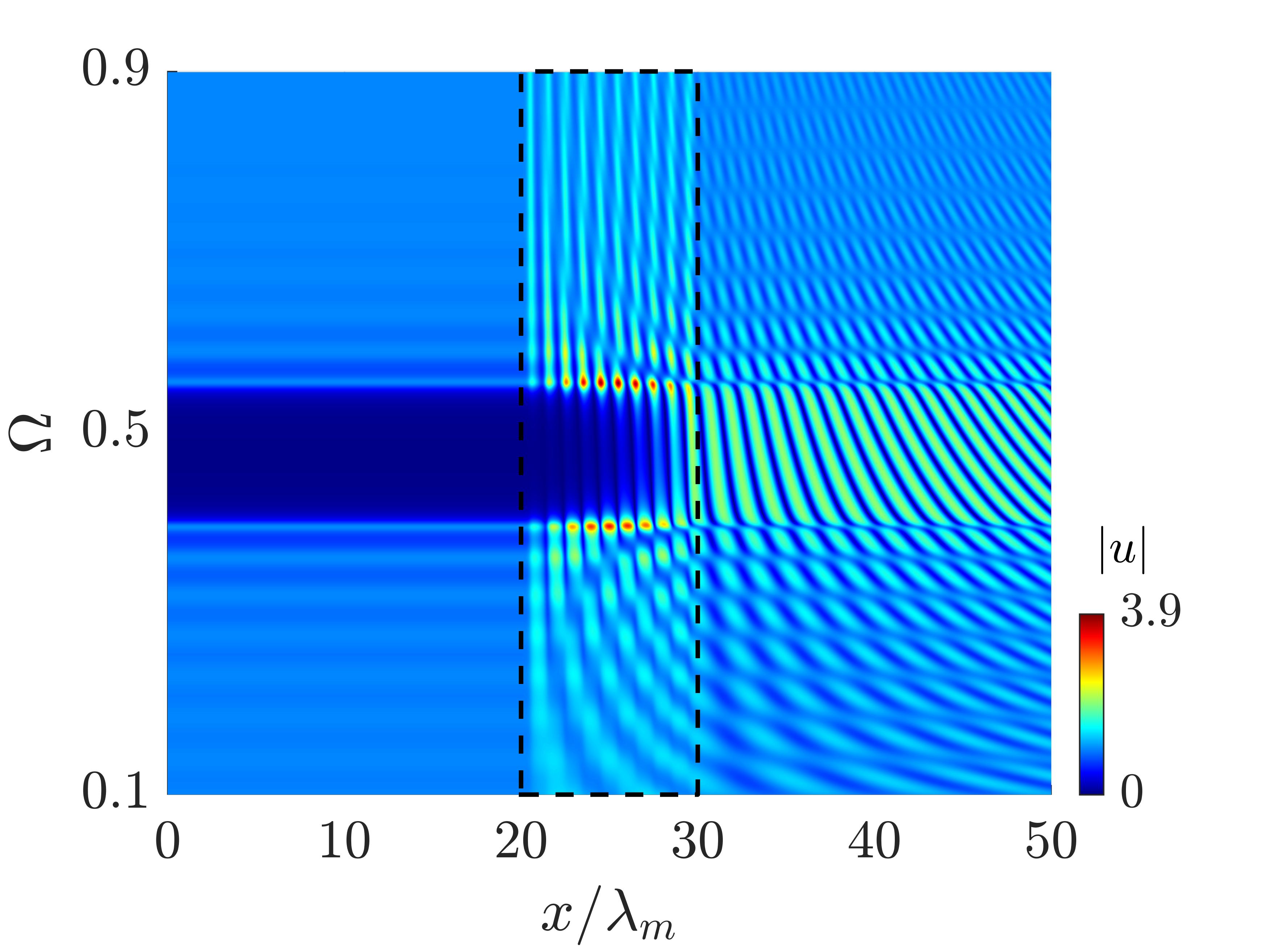}}\\
	\subfigure[]{\includegraphics[width=0.4\textwidth]{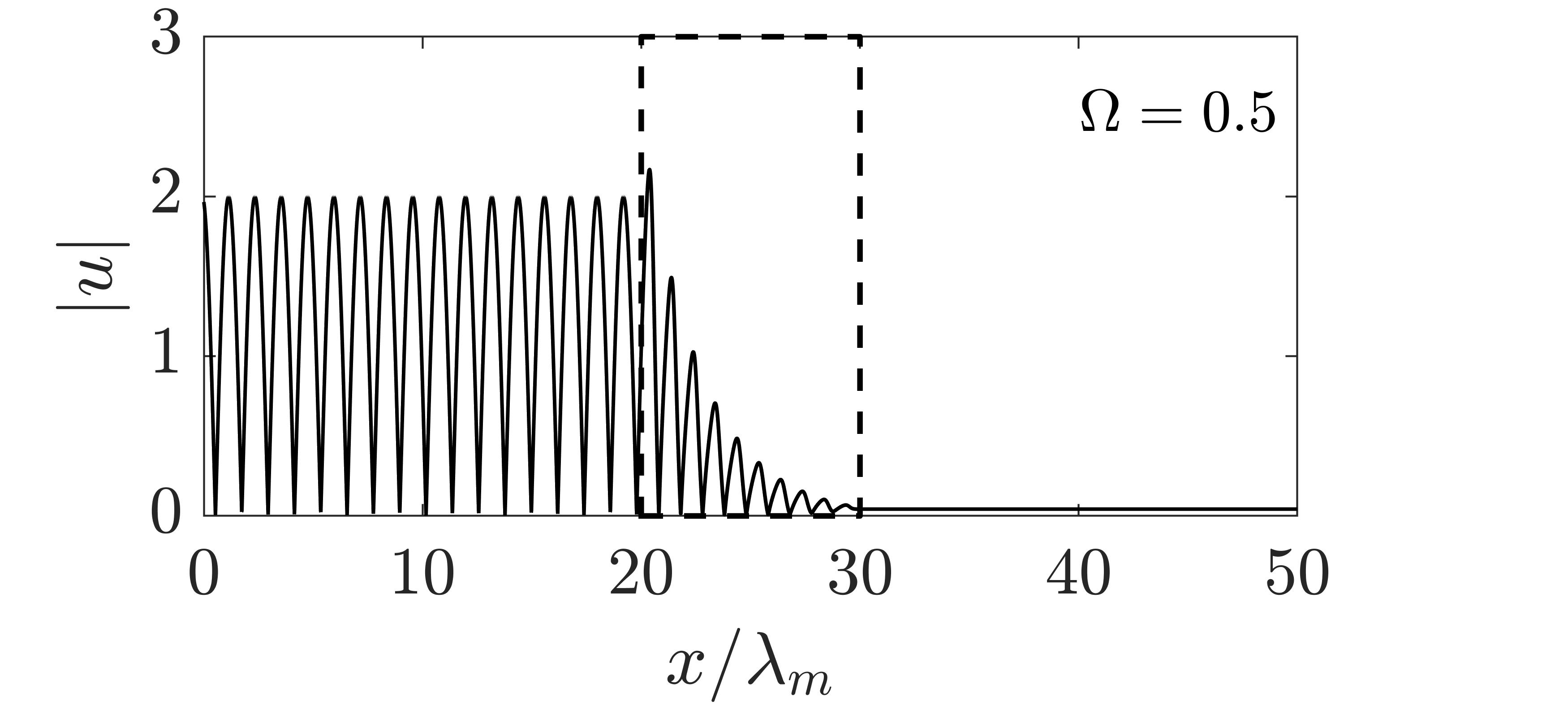}}
	\subfigure[]{\includegraphics[width=0.4\textwidth]{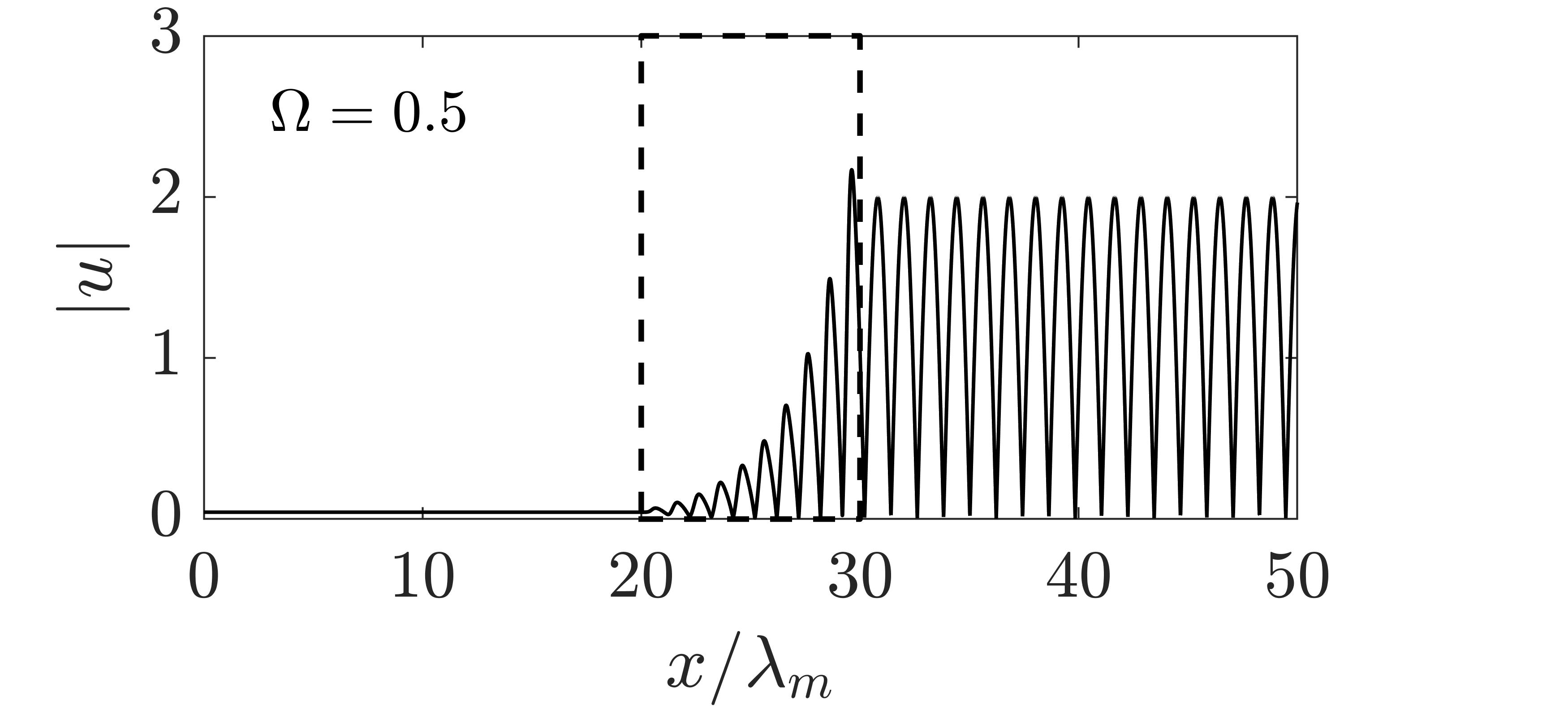}}\\	\subfigure[]{\includegraphics[width=0.4\textwidth]{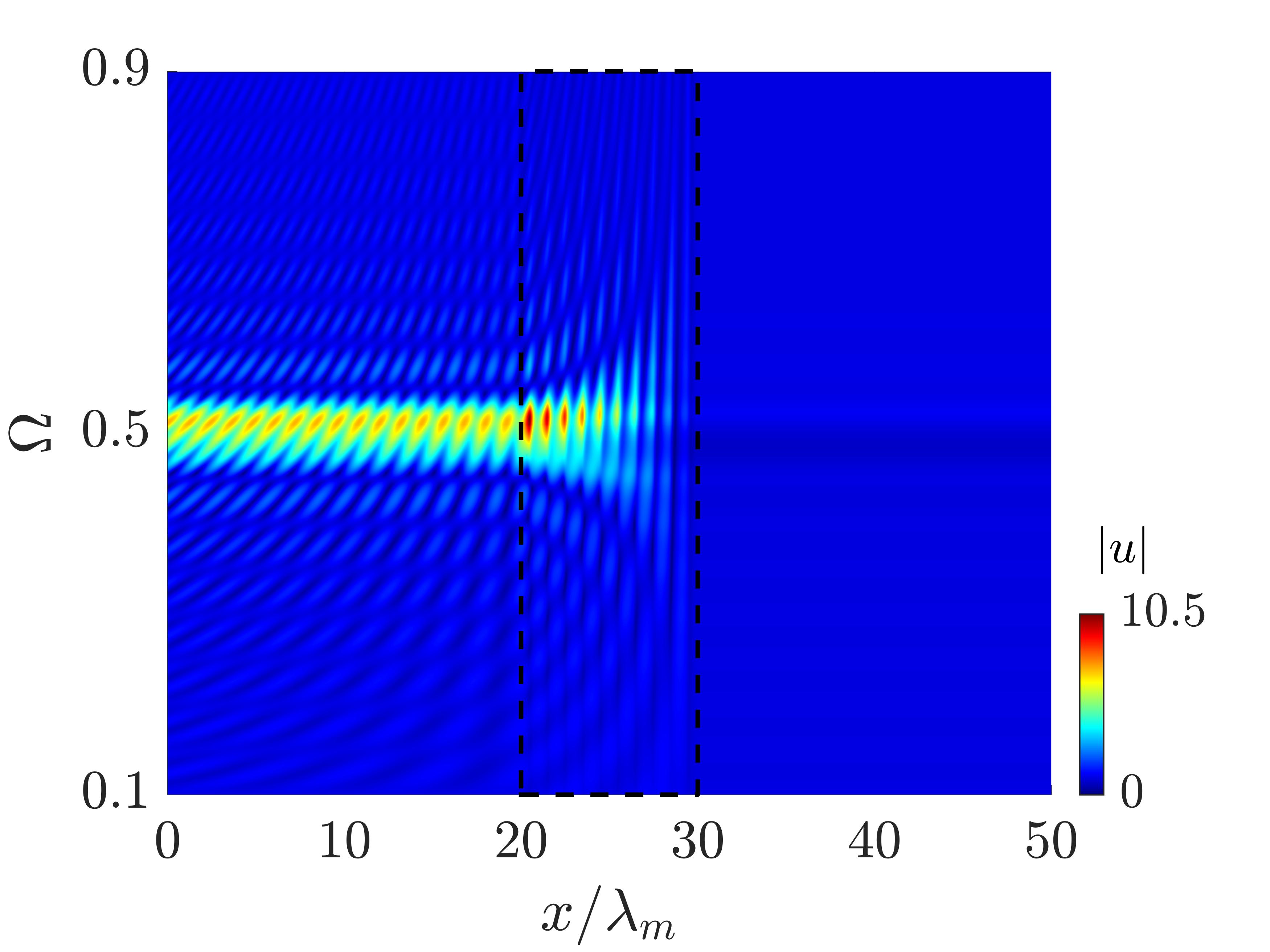}}
	\subfigure[]{\includegraphics[width=0.4\textwidth]{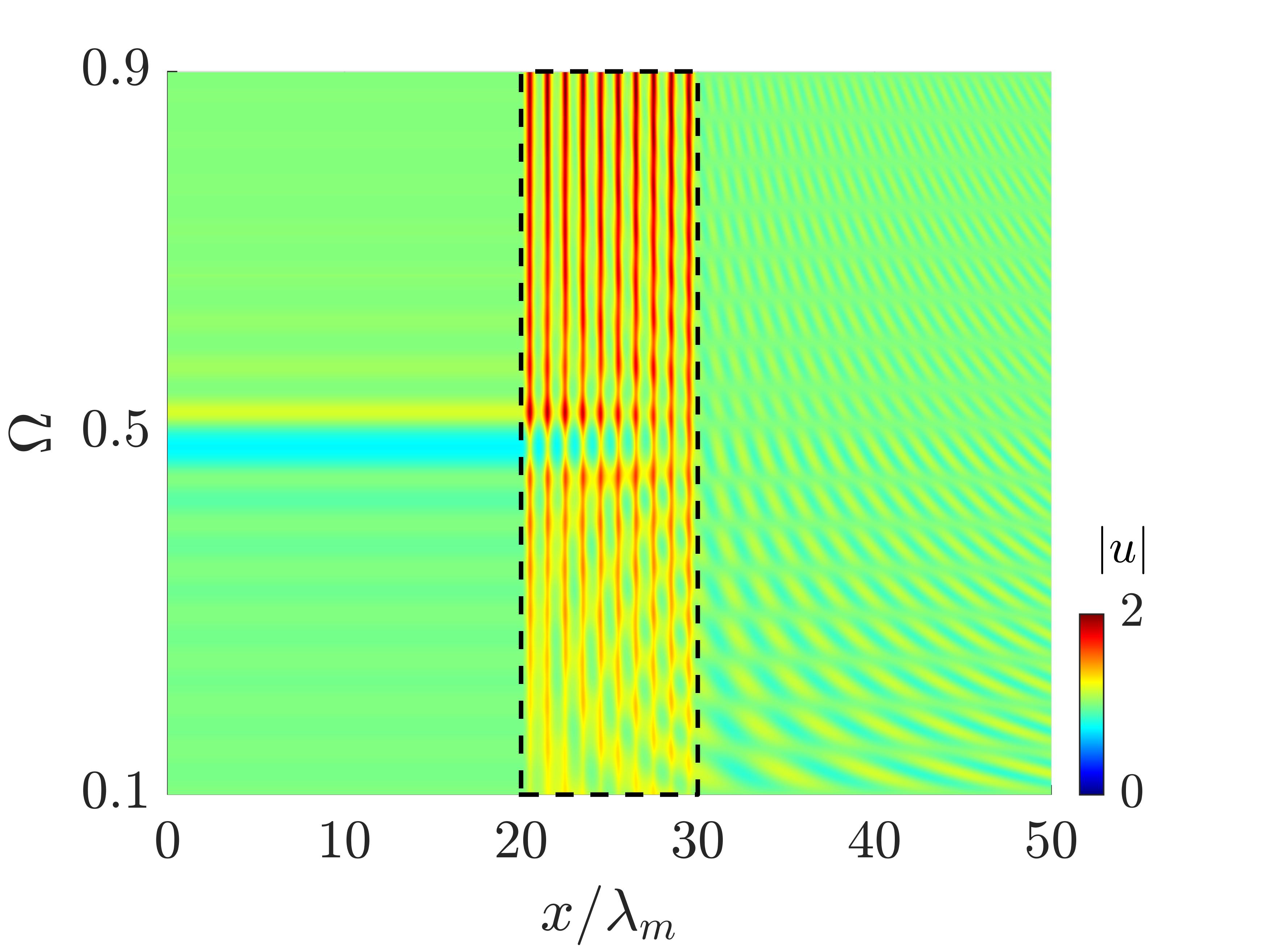}}\\
    \subfigure[]{\includegraphics[width=0.4\textwidth]{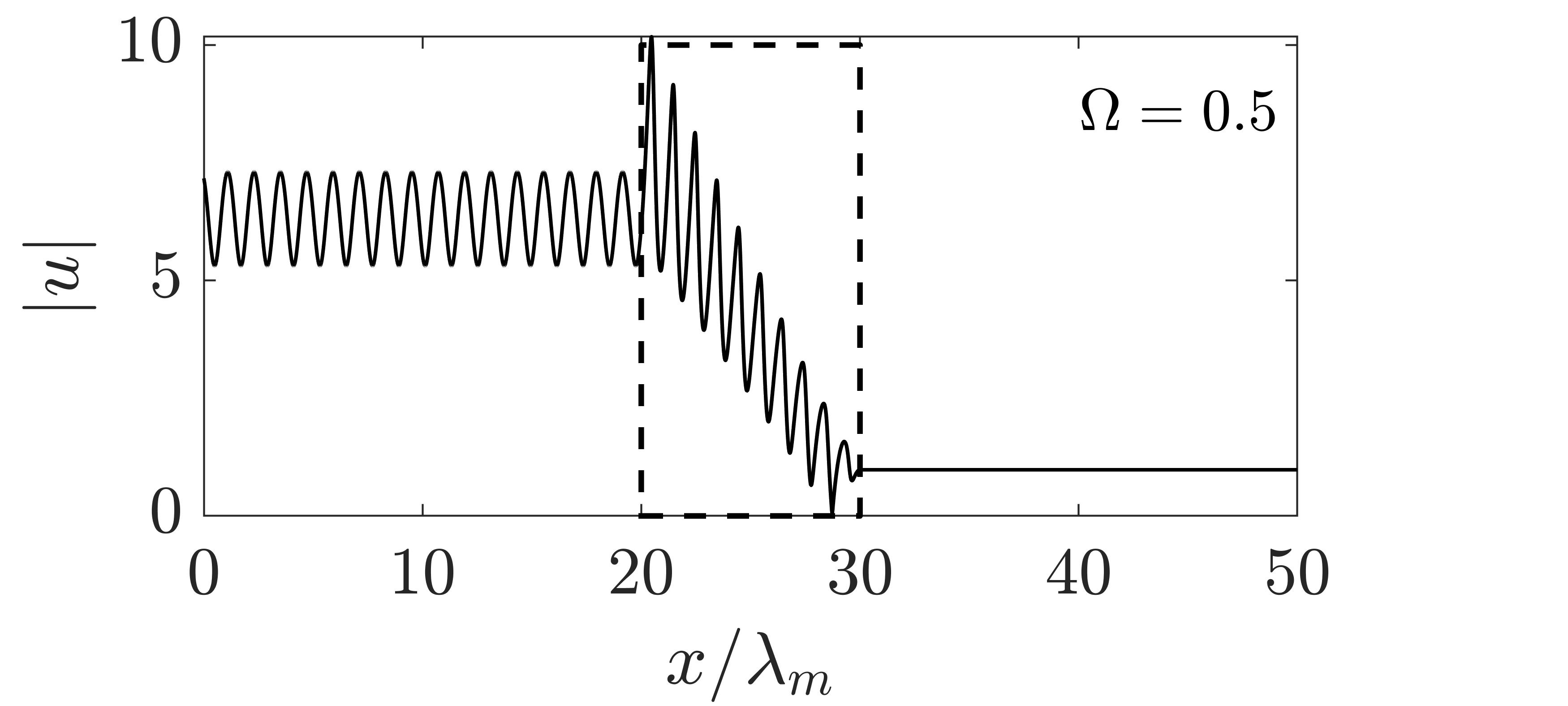}}
	\subfigure[]{\includegraphics[width=0.4\textwidth]{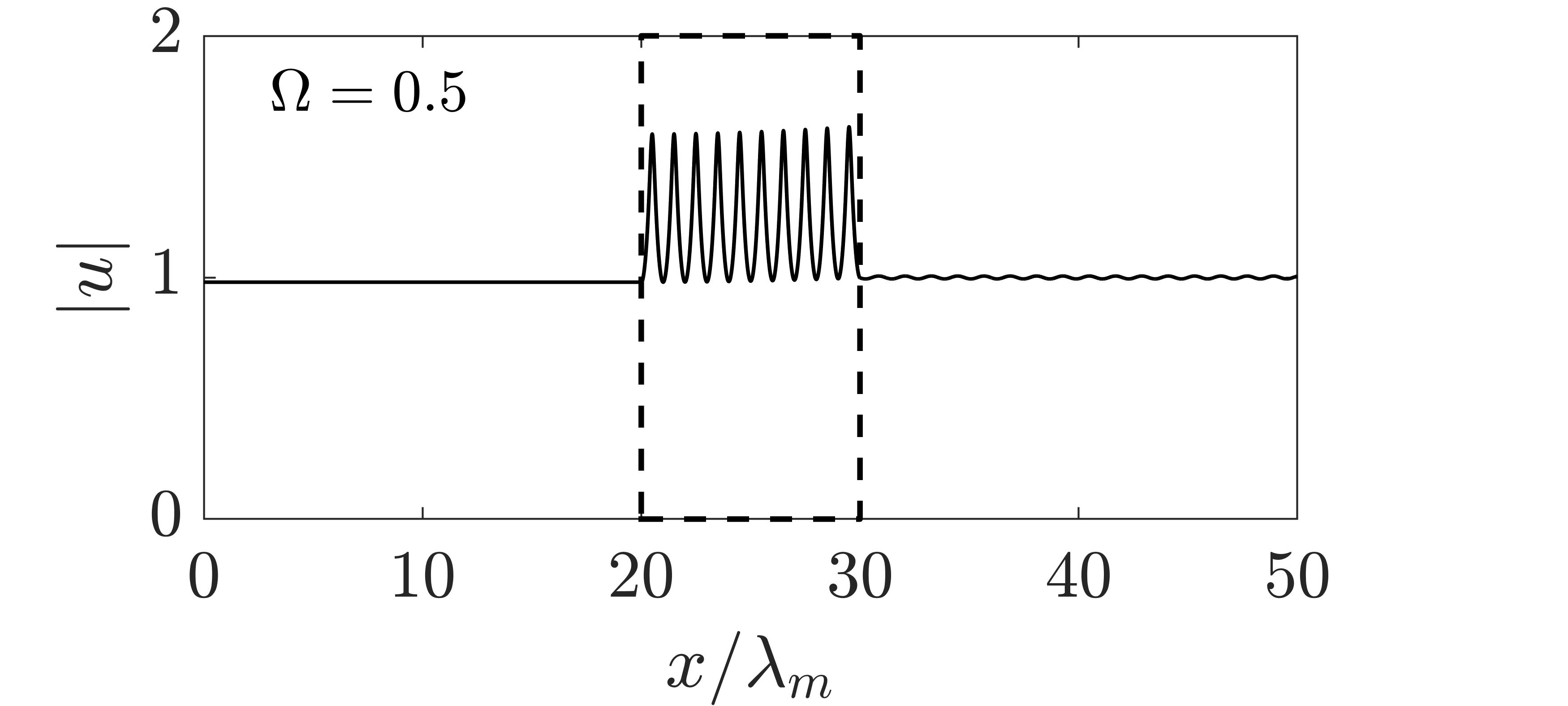}}
	\caption{Displacement field $|u\left(x,\Omega\right)|$ upon varying frequency $\Omega$ and with $\gamma=0$ for a wave impinging (a) from the left and (b) from the right. (c,d) Same displacement fields for $\Omega=0.5$. (e) Displacement field $|u\left(x,\Omega\right)|$ upon varying frequency $\Omega$ and with $\gamma=0.5$ for a wave impinging (e) from the left and (f) from the right. The response is clearly asymmetric. (g-h) Same response but limited to $\Omega=0.5$.}
	\label{Fig3}    
\end{figure*}
Some considerations follow: (i) when the gain/loss interactions are not active, the reflections are almost unitary and the transmissions drop to zero. Furthermore, $T_{12}=T_{21}$, $R_{11}=R_{22}$ and, hence, the medium is reciprocal.
(ii) It is observed that, in correspondence of the exceptional points, the reflection from the right side of the slab drops to zero ($R_{22}=0$), while the transmission is unitary ($T_{21}=1$). In contrast, wave propagation from the left to the right exhibits reflection with amplification ($R_{11}>>1$), which is accompanied by unitary transmission ($T_{12}=1$) as per right-to-left propagating waves and similarly to prior works on this matter \cite{wu2019asymmetric,fleury2015invisible}. This confirms the asymmetric scattering capabilities of the slab that, as mentioned, are justified by the emergence of asymmetric wave modes at the non-Hermitian degeneracy.
The analysis of the scattering matrix is concluded in the supplementary material \cite{SM}, where the scattering coefficients are reported upon varying $\gamma$.\\
Now, given the relationship between the amplitude coefficients, the transmitted and reflected wave fields generated by left and right impinging waves are hereafter studied, with emphasis on the behavior at the EP. It is initially assumed that a wave with unitary amplitude propagates from left to right, i.e. $A_0=1$ and $G_0=0$. As such, the scattered wave amplitudes $B_0$ and $F_0$ are due to Eq. \ref{eq:scattering} which, when combined to the waves in Eq. \ref{eq:06} and Eq. \ref{eq:07}, give the displacement field outside the modulated slab. $C_0$ and $D_0$ are instead evaluated through Eq. \ref{eq:09} and are employed to reconstruct the response between $x_1$ and $x_2$. The same procedure is applied to right impinging waves, i.e. $A_0=0$ and $G_0=1$. The response of the slab for the two excitation conditions is illustrated in Figure \ref{Fig3}(a-d) in terms of wave amplitude $|u\left(x,\Omega\right)|$ and, initially, for $\gamma=0$. Figure \ref{Fig3}(a) is relative to a wave that propagates from the left to the right and, to ease visualization, the response for $\Omega=0.5$ is reported in Figure \ref{Fig3}(c). Vice-versa Figure \ref{Fig3}(b,d) illustrate the behavior for a right-impinging wave. 
\begin{figure*}[t!]
	\centering
	\subfigure[]{\includegraphics[width=0.35\textwidth]{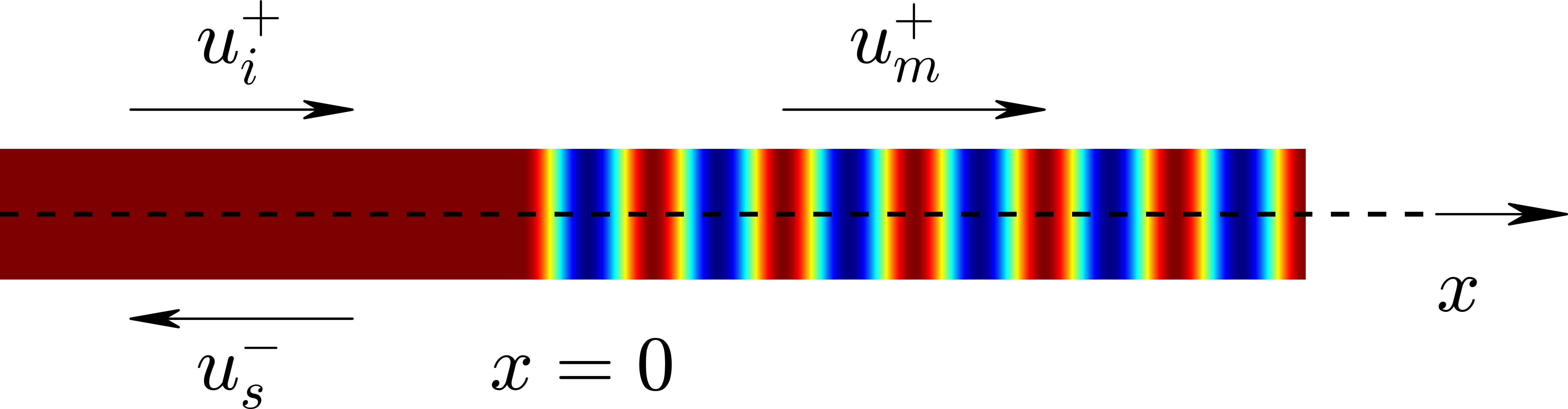}}\hspace{2.5cm}
	\subfigure[]{\includegraphics[width=0.35\textwidth]{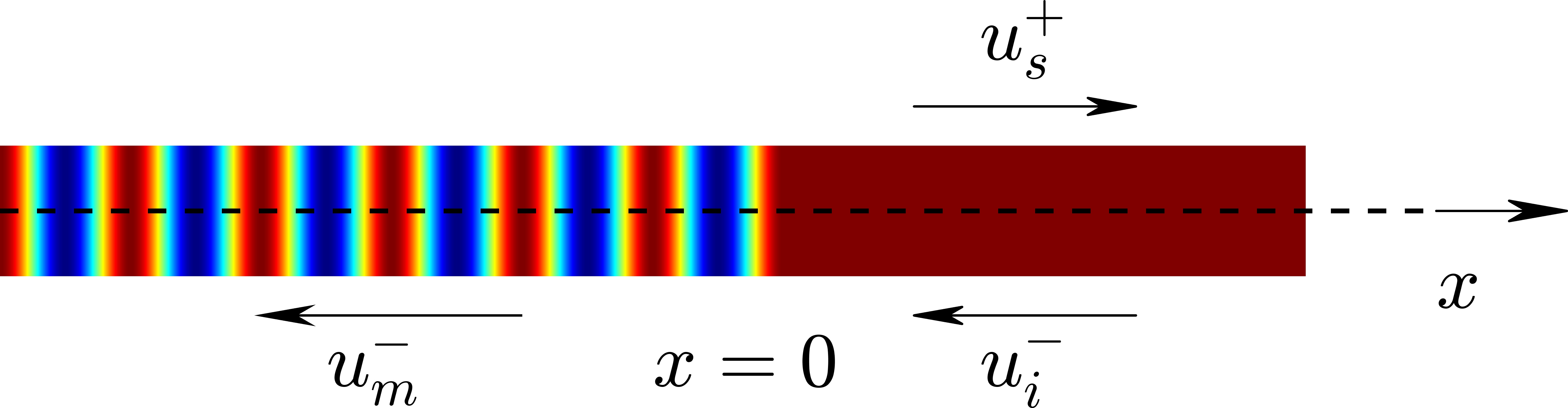}}\\
	\subfigure[]{\includegraphics[width=0.48\textwidth]{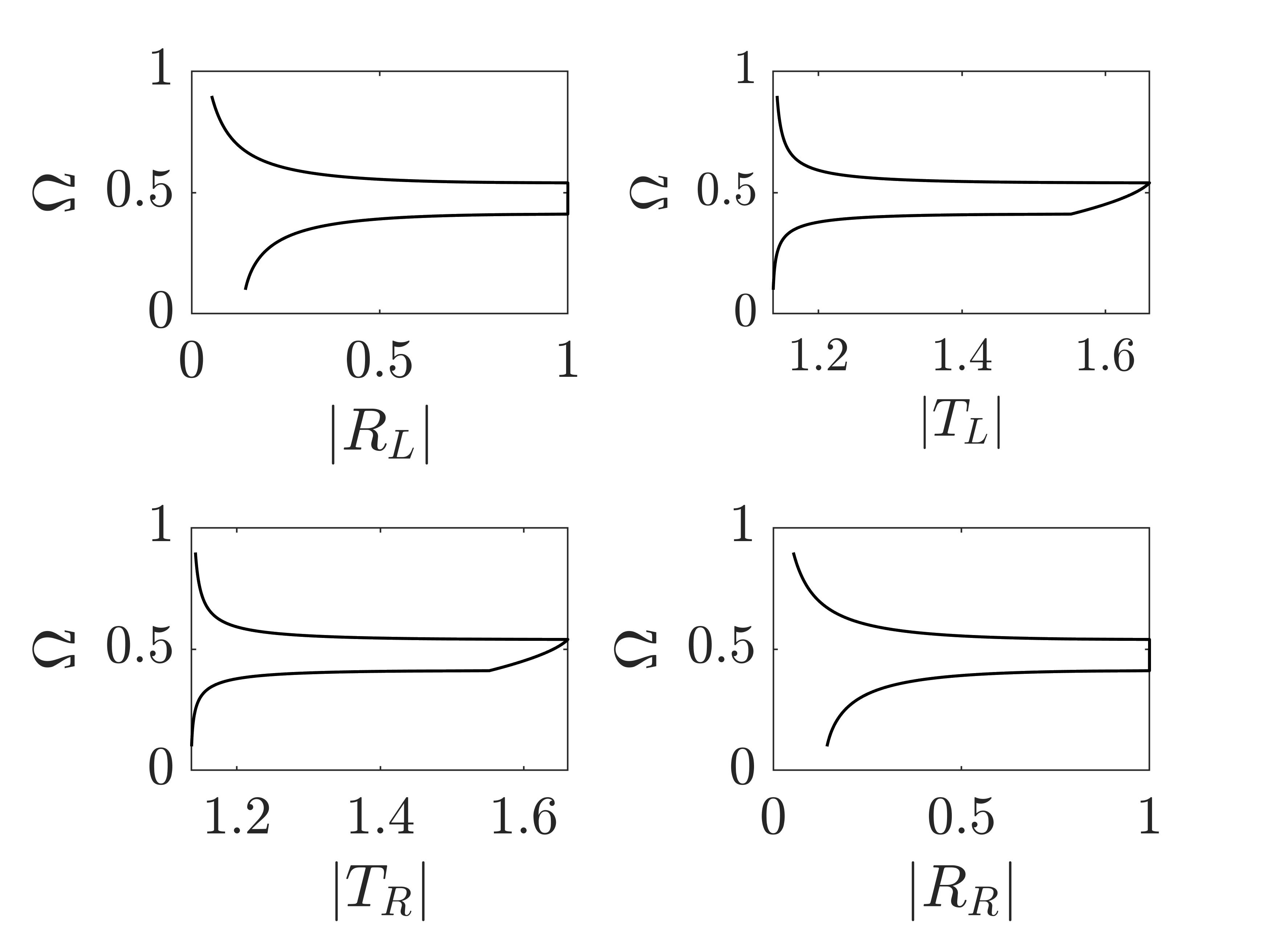}}
	\subfigure[]{\includegraphics[width=0.48\textwidth]{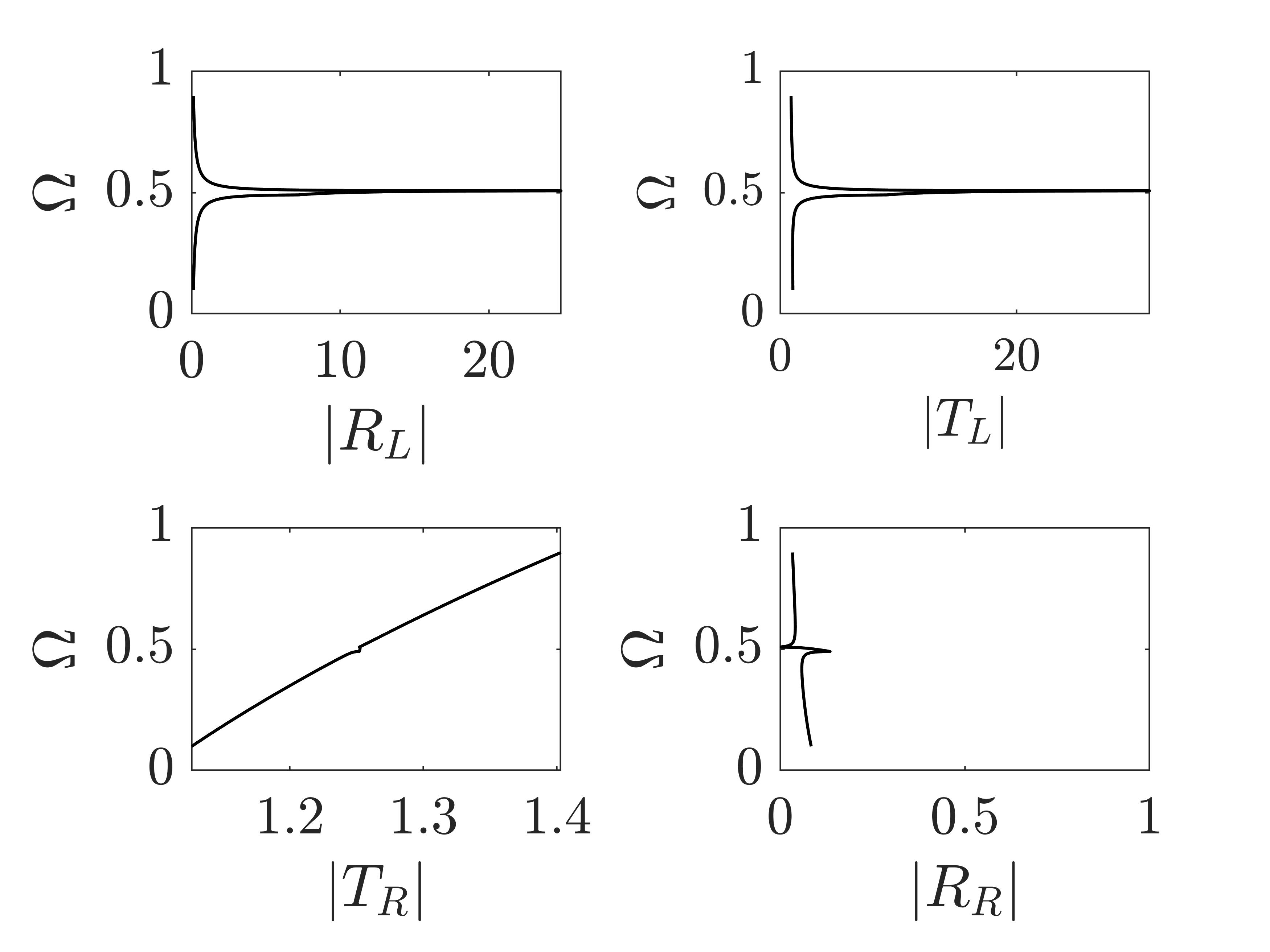}}
	\caption{Schematic of the interface between a semi-infinite homogeneous domain (red) and a modulated semi-infinite medium (colored part). Modulation is on the (a) left or on the (b) right side of the interface. The incident and scattered waves are displayed with arrows. (b) Scattering coefficients of the interface for (c) $\gamma=0$ and (d) $\gamma=\alpha=0.5$. }
	\label{Fig5}    
\end{figure*}
It is straightforward to conclude that, since the diagrams for the mirror-symmetric inputs are mirror-symmetric about the center of the slab, the medium is reciprocal. In addition to that, the slab exhibits exponential attenuation for bandgap frequencies, regardless the direction of the impinging wave. Consider now the case $\gamma=\alpha$, which is illustrated in Figure \ref{Fig3}(e-h). When a plane wave with frequency in the neighborhood of $\Omega=0.5$ impinges from the left, there is reflection with a strong amplification factor, see Figure \ref{Fig3}(e,g). Such a strong selective amplification is accompanied by unitary transmission. In contrast, when a wave impinges from the right, see Figure \ref{Fig3}(f,h) there is no reflection and the wave is transmitted unaltered. This blueprint of nonreciprocity is justified by the directional wave modes $\hat{u}_{0,p}^\pm$ which take part on the scattering process described in Eq. \ref{eq:scattering} and yield directional scattering coefficients. Similar results are obtained through a modulated slab with a single unit of dimension $\lambda_m$, and reported in the supplementary material \cite{SM}, where the benefits of periodic configurations are also discussed.\\
The paper is concluded with the analysis of the two interfaces, to provide additional insight on the role of the wave modes on the directional scattering capabilities. Consider the lone interfaces at $x_1$ and $x_2$, which now consists of two semi-infinite media and, for simplicity, a new reference system is added in correspondence of the interface, see the schematic in Figure \ref{Fig5}(a-b). The homogeneous medium is located at the left (right) of the interface and rigidly connected to the modulated slab. Since the media are now semi-infinite, a left or right incident waves of amplitude $A_0$ or $G_0$ are capable of exciting either the $u_m^+$ or the $u_m^-$ wave modes, and the reflections are dictated by new equilibrium conditions. For the left interface:
\begin{equation}
\begin{split}
    &u_i^+\left(0\right)+u_s^-\left(0\right)=u_m^+\left(0\right)\\[5pt]
    &E_0\left(1+\alpha\right)\left(u_i^+\left(0\right)+u_s^-\left(0\right)\right)_{,x}= E\left(0\right)\left(u_m^+\left(0\right)\right)_{,x}
\end{split}
\label{eq:12}
\end{equation}
for the right interface:
\begin{equation}
\begin{split}
    &u_m^-\left(0\right)=u_i^-\left(0\right)+u_s^+\left(0\right)\\[5pt]
    &E\left(0\right)\left(u_m^+\left(0\right)\right)_{,x}=\\
    &\hspace{2.5cm}E_0\left(1+\alpha\right)\left(u_i^-\left(0\right)+u_s^+\left(0\right)\right)_{,x}
\end{split}
\end{equation}
These equations are combined to get to an expression for the reflected and transmitted wave amplitudes. For the left interface:
\begingroup\makeatletter\def\f@size{9}\check@mathfonts
\begin{equation}
\begin{split}
    T_{L}=\frac{C_0}{A_0}=\frac{2}{\displaystyle\sum_{p=-P}^{P}\hat{u}_{p,0}^++\displaystyle\frac{E_0(0)}{E_0\left(1+\alpha\right)}\frac{c_{0,\alpha}}{\omega}\sum_{p=-P}^{P}\hat{u}_{p,0}^+\left(\kappa_0^++p\kappa_m\right)}\\[5pt]
    R_{L}=\frac{B_0}{A_0}=\frac{\displaystyle\sum_{p=-P}^{P}\hat{u}_{p,0}^+-\displaystyle\frac{E_0(0)}{E_0\left(1+\alpha\right)}\frac{c_{0,\alpha}}{\omega}\sum_{p=-P}^{P}\hat{u}_{p,0}^+\left(\kappa_0^++p\kappa_m\right)}{\displaystyle\sum_{p=-P}^{P}\hat{u}_{p,0}^++\displaystyle\frac{E_0(0)}{E_0\left(1+\alpha\right)}\frac{c_{0,\alpha}}{\omega}\sum_{p=-P}^{P}\hat{u}_{p,0}^+\left(\kappa_0^++p\kappa_m\right)}
\end{split}    
\label{eq:14}
\end{equation}
\endgroup
and for a right interface, when a wave impinges from the right:
\begingroup\makeatletter\def\f@size{9}\check@mathfonts
\begin{equation}
\begin{split}
    T_{R}=\frac{D_0}{G_0}=\frac{2}{\displaystyle\sum_{p=-P}^{P}\hat{u}_{p,0}^--\displaystyle\frac{E_0(0)}{E_0\left(1+\alpha\right)}\frac{c_{0,\alpha}}{\omega}\sum_{p=-P}^{P}\hat{u}_{p,0}^-\left(\kappa_0^-+p\kappa_m\right)}\\[5pt]
    R_{R}=\frac{F_0}{G_0}=\frac{\displaystyle\sum_{p=-P}^{P}\hat{u}_{p,0}^-+\displaystyle\frac{E_0(0)}{E_0\left(1+\alpha\right)}\frac{c_{0,\alpha}}{\omega}\sum_{p=-P}^{P}\hat{u}_{p,0}^-\left(\kappa_0^-+p\kappa_m\right)}{\displaystyle\sum_{p=-P}^{P}\hat{u}_{p,0}^--\displaystyle\frac{E_0(0)}{E_0\left(1+\alpha\right)}\frac{c_{0,\alpha}}{\omega}\sum_{p=-P}^{P}\hat{u}_{p,0}^-\left(\kappa_0^-+p\kappa_m\right)}
\end{split}    
\label{eq:15}
\end{equation}
\endgroup
it is straightforward to conclude that waves impinging from different directions (left $(\cdot)_L$ or right $(\cdot)_R$), capable of exciting different wave modes $\hat{u}_{p,0}^+\neq\hat{u}_{p,0}^-$, generate different reflection ($R_{L,R}$) and transmission ($T_{L,R}$) coefficients at the interface. This is evident in Eq. \ref{eq:14}-\ref{eq:15} and illustrated in Figure \ref{Fig5}(c-d) where $T_{L}$, $R_{L}$, $T_{R}$, and $R_{R}$ are displayed upon varying frequency. The results for $\gamma=0$ and $\gamma=\alpha$ are shown in Figure \ref{Fig5}(c) and Figure \ref{Fig5}(d), respectively. While for completeness, different $\gamma$ values are provided in the supplementary material \cite{SM}.
As expected, the curves for $\gamma=0$ are identical for waves impinging from the left or from the right. In contrast, complex modulations with $\gamma=\alpha$ display a sharp peak if waves impinge from the left. Waves impinging from the right are not reflected at all, which further confirms the asymmetric scattering capabilities of the modulated slab. 

\section*{Conclusions}
In this manuscript, the plane wave expansion method (PWEM) and the scattering matrix method (SMM) are employed to study asymmetric scattering of longitudinal waves in a $\mathcal{PT}$-symmetric modulated beam with complex elasticity. It is shown that such a system support either frequency gaps and $\kappa$-gaps depending on the amount of complex stiffness modulation. The transition among them is characterized by dispersion branches that coalesce into an exceptional point, which separates the two phases. This condition triggers a rich scattering behavior, which is studied in the paper with emphasis on the role of the wave modes. In analogy with prior works, it is shown that waves impinging from opposite directions can be either transmitted with no reflection (right to left), or reflected with strong amplification and transmitted with unitary amplitude (left to right). The directional behavior of the slab is justified by the asymmetric wave modes supported by the slab, which are selectively excited at frequencies in the neighborhood of the EP by left or right impinging waves. This work can be extended to elastic waveguides supporting different wave modes (e.g. torsional, flexural) and the observations made in the paper are key for the design of elastic systems with asymmetric capabilities that can be functional for signal processing, vibration control in structures, and selective amplification.\\

\begin{acknowledgments}
E.R. wishes to thank M. Rosa for useful discussion. 
\end{acknowledgments}

\appendix

\nocite{*}

\bibliography{apssamp}
 
\end{document}